\documentclass[final,5p,times,twocolumn]{elsarticle}

\usepackage{graphicx}
\usepackage{amssymb}
\usepackage{amsthm}
\usepackage{amsmath}
\usepackage{lineno}
\usepackage{subfigure}

\usepackage{upgreek}

\biboptions{sort&compress}

\usepackage{booktabs}

\usepackage{mathptmx}
\usepackage{lscape}
\def\imagebox#1#2{\vtop to #1{\null\hbox{#2}\vfill}}
\usepackage{array}
\usepackage{droidsans}
\usepackage{charter}
\usepackage[usenames,dvipsnames]{xcolor}
\usepackage{setspace}
\usepackage[compact]{titlesec}
\usepackage{hyperref}
\usepackage{chemformula}
\usepackage[detect-none]{siunitx}
\usepackage{booktabs}
\usepackage{caption}
\usepackage{epstopdf}

\journal{Journal of molecular liquids}

\begin{document}

\begin{frontmatter}


\title{Behaviour of a magnetic nanogel in a shear flow}



\author[First]{Ivan S. Novikau$^\ast$}
\ead{ivan.novikau@univie.ac.at}
\author[Second]{Ekaterina V. Novak}
\author[Second]{Elena S. Pyanzina}
\author[First,Second,Third]{Sofia S. Kantorovich}

\address[First]{Faculty of Physics, University of Vienna, Kollingasse 14-16, 1090 Vienna, Austria}
\address[Second]{Institute of Mathematics and Natural Sciences, Ural Federal University, Lenin av. 51, 620000, Ekaterinburg, Russia}
\address[Third]{Research Platform MMM Mathematics-Magnetism-Material, University of Vienna, Oskar-Morgenstern-Platz 1, 1090 Vienna, Austria}

\begin{abstract}
Magnetic nanogels (MNG) -- soft colloids made of polymer matrix with embedded in it magnetic nanoparticles (MNPs) -- are promising  magneto-controllable drug carriers. In order to develop this potential, one needs to clearly understand the relationship between nanogel magnetic properties and its behaviour in a hydrodynamic flow. Considering the size of the MNG and typical time and velocity scales involved in their nanofluidics, experimental characterisation of the system is challenging. In this work, we perform  molecular dynamics (MD) simulations combined with the Lattice-Boltzmann (LB) scheme aiming at describing the impact of the shear rate ($\dot{\gamma}$) on the shape, magnetic structure and motion of an MNG. We find that in a shear flow the centre of mass of a MNG tends to be in the centre of a channel and to move preserving the distance to both walls. The MNG monomers along with translation are involved in two more types of motion, they rotate around the centre of mass and oscillate with respect to the latter. It results in synchronised tumbling and wobbling of the whole MNG accompanied by its volume oscillates. The fact the a MNG is a highly compressible and permeable for the carrier liquid object makes its behaviour different from that predicted by classical Taylor-type models. We show that the frequency of volume oscillations and rotations are identical and are growing function of the shear rate. We find that the stronger magnetic interactions in the MNG are, the higher is the frequency of this complex oscillatory motion, and the lower is its amplitude. Finally, we show that the oscillations of the volume lead to the periodic changes in MNG magnetic energy.
\end{abstract}


\end{frontmatter}


\section{Introduction}

More than 60 years passed since the term ``microgel'' was first applied to describe soft colloidal particles made of crosslinked polymers  \cite{Baker1949,2007-aleman-pac,2011-fernandez-nieves-bk, 2017-hamzah-jpr}. Nowadays, one usually distinguishes between micro- and nanogels depending on their characteristic size.  Fast development in synthesis techniques \cite{2011-fernandez-nieves-bk,Bonham2014,2016-mavila-chrv,2017-hamzah-jpr, 2016-galia-eccm,zoe21a} in the last decade resulted in nano- and microgels responsive to various stimuli such as pH \cite{Hoare2004,Yin2008}, temperature \cite{Hoare2004,Mohanty2015,Backes2017a}, electromagnetic radiation \cite{Gorelikov2004}, ionic strength or electric fields \cite{Mohanty2015, Colla2018}. Among typical responses are  swelling/collapse transitions or particular rheological properties \cite{2018-doukas-sm,2017-foglino-prl,niezabitowska20a,scotti20a,lopez21a}.  

The ability of micro- and nanogels to respond to external stimuli puts them in the class of smart materials with a large potential in biomedicine and technology \cite{Das2006, 2011-thorne-cps, Hu2012, Bonham2014,Agrawal2018, wang20a,cornejobravo21a,vashist20a}. Among prominent examples are  responsive coatings \cite{Raquois1995,Saatweber1996}, chemical sensors and biosensing probes \cite{RETAMA2003,Guo2005,2014-sigolaeva-bmm,Aliberti2017} or 3D printing \cite{highley19a}. Microgels are employed in tissue engineering \cite{Pepe2017},  manipulation and template-based synthesis of solid nanoparticles \cite{Bayliss2011, Zhang2004a}, as well as in water management and oil or pollutant recovery \cite{Son2016,Alhuraishawy2018,PU2019}. The size range of nanogels enables efficient drug delivery as they can cross biological barriers   \cite{LOPEZ2005,2008-kwon-oh-pps,Vinogradov2010,2011-malmsten-ch,Sivakumaran2011,Pepe2017,Schimka2017,toprakcioglueaay20a, sawada20a}.

To broaden the spectra of bio-compatible control stimuli and make the drug delivery even more targeted, it was proposed to embed magnetic nanoparticles into the polymer network \cite{2004-menager-pol,2015-backes-jpcb,mandal19a,Witt2019,sung20a,cao20a,gao20a,biglione20a}, thus creating magnetic micro- and nanogels, abbreviated to MMG and MNG respectively. The idea to combine magnetic nanoparticles with polymer matrix had been previously tested on bulk systems -- macroscopic magnetic gels \cite{1995-shiga, 2000-zrinyi, 2010-reinicke, 2011-frickel, 2011-messing, 2013-ilg, 2013-xu, 2015-roeder,suarezfernandez20a,veloso21a}. Such macroscopic magnetic gels were shown to reversibly and efficiently change their mechanic and optical properties on the application of an external magnetic field of a moderate strength. 

Considering the aforementioned perspectives that nano- and microgels open up in small scale fluidics, a plethora of approaches, experimental and theoretical, were put forward to study their structural, mechanical, and rheological properties \cite{2006-hoare-jpcb,pellet16a, 2018-backes-pol, 2019-witte-sm,Kobayashi2014,Ghavami2016,Kobayashi2016,Ahualli2017,Ghavami2017,Kobayashi2017,Gnan2017,Hofzumahaus2018, 2019-martin-molina-jml-rev, 2020-scotti-flow}. In particular, computer simulations with a realistic representation of the polymer network shed  light on mechanical response  of microgels \cite{Gnan2017,Moreno2018,2019-rovigatti-sm-rev,ninarello2019modeling,camerin2020microgels,rovigatti2019connecting} as well as on their electrostatics \cite{del2020charge}. Aiming at avoiding lattice-based polymer networks, we developed a coarse-grained simulation model of magnetic and nonmagnetic nanogels with a non regular internal structure \cite{Minina2018,2019-minina-jml} that qualitatively reproduce the structural features of nanogels obtained by electrochemically or photonically induced crosslinking of polymer precursors confined in emulsion nanodroplets \cite{2016-galia-eccm,2016-mavila-chrv}.  We employed the aforementioned model  to investigate the equilibrium structural properties of a single MNG \cite{Minina2018}, as well as MNG suspensions \cite{2019-novikau-icmf} in absence of an applied external field. It was found that inside a single MNG, magnetic nanoparticles tend to form small clusters whose shape is largely affected by the amount of crosslinkers. In suspension, MNGs can self-assemble by forming magnetic nanoparticle bridges between themselves, but only if their concentration is sufficiently high and the magnetic interparticle interaction is strong.  While studying the magnetic response of MNG suspension to external fields \cite{novikau2020influence}, we discovered that  for the same bulk magnetic content, the magnetic particles embedded in MNGs are more susceptible to an applied field than an ensemble of their free counterparts. This is the result of locally concentrated regions inside the gels. However, the  elastic constraints make the susceptibility lower than that of the ensemble of free particles with the volume fraction of nanoparticles inside the gel. 

The majority of micro- and nanogel applications rely on the rheology of their diluted or concentrated suspensions. Thus, the behaviour of nonmagnetic nano- and microgels in the flow was actively studied in the last years revealing the ability of the flow to change structure and mechanics of these systems \cite{bhattacharjee18a,torres18emulsion,camerin2018modelling,fromanek19a,formanek21a} opening up new challenges in the field.  Even less understood is the rheology of magnetic nanogles \cite{mandal19a}. It makes it particularly relevant to elucidate the effects of combined hydrodynamic and long-range magnetic interactions on the structure and properties of MNGs in the flow. In this study we make the first step in this direction and investigate a single MNG in the shear flow using combined Molecular Dynamics and Lattice-Boltzmann approach. 

The paper is organised as follows. In section \ref{sec:sim-m} we discuss the main ingredients of the simulation model and the protocol. Aiming at detail understanding of the interplay between shear and magnetic interparticle interactions we first analyse the shape of the MNG as a function of the shear rate in section \ref{sec:shape-eval} and show the periodic change in MNG overall shape. Next, in section \ref{sec:matr}, we investigate the motion of the polymer matrix with respect to the MNG centre of mass, revealing a periodic tumbling. In section \ref{sec:magn} we closely follow the behaviour of the magnetic nanoparticles inside the gel and show how even a very weak flow prevents the formation of large magnetic particle clusters observed in the equilibrium conditions. It follows by section \ref{sec:positioning}, where we compare obtained results with the similar findings of other studies. Conclusions and outlook are provided in section \ref{sec:con}. Appendix A discusses the forces acting on the MNG from the walls of the channel that we used to justify the choice of the system parameters.

\section{Simulation Approach}\label{sec:sim-m}

\begin{figure}[h]
\centering
  \includegraphics[width=0.49\textwidth]{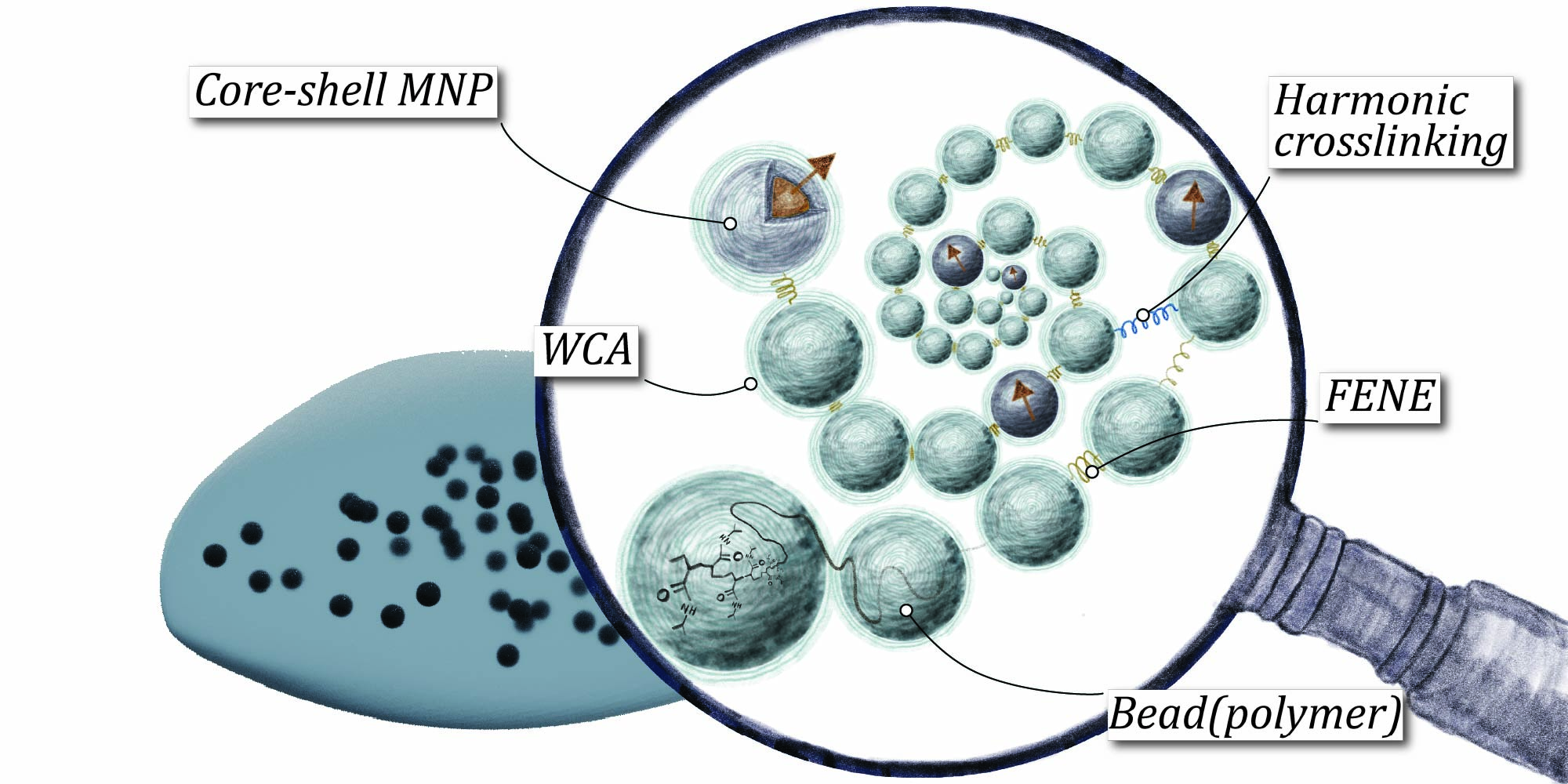}
  \caption{Model. In the back is the MNG. Magnetic particles are shown in black, polymer is replaced by its convex hull. The zoom-in explains the main interactions.}
  \label{fig:model1}
\end{figure}

Our MNG consists of crosslinked polymer chains represented by a classical bead-spring model \cite{Kremer1990} with some beads being randomly replaced by magnetic particles \cite{Minina2018, 2019-minina-jml}.  The schematic of such a soft magnetic complex colloid is shown in Fig. \ref{fig:model1}. Here, the MNG is shown in the back as a convex hull, magnetic particles are in black. Increasing the resolution, as shown in the inset, one can see that the polymer beads, whose sizes are those of characteristic blobs, and magnetic particles have an excluded volume and their centres are interconnected by springs. We use two different types of magnetic nanogels, depending on the magnetic particles they contain: cobalt ferrite (\ch{CoFe2O4}) or cobalt (\ch{Co}). In both cases those particles are considered to be sterically stabilised with the shell of approximately 2 nm thickness. This is reflected in the choice of particle densities as explained below.

\subsection{Dimensionless Units}
\begin{figure*}[h]
\centering
  \includegraphics[width=0.95\textwidth]{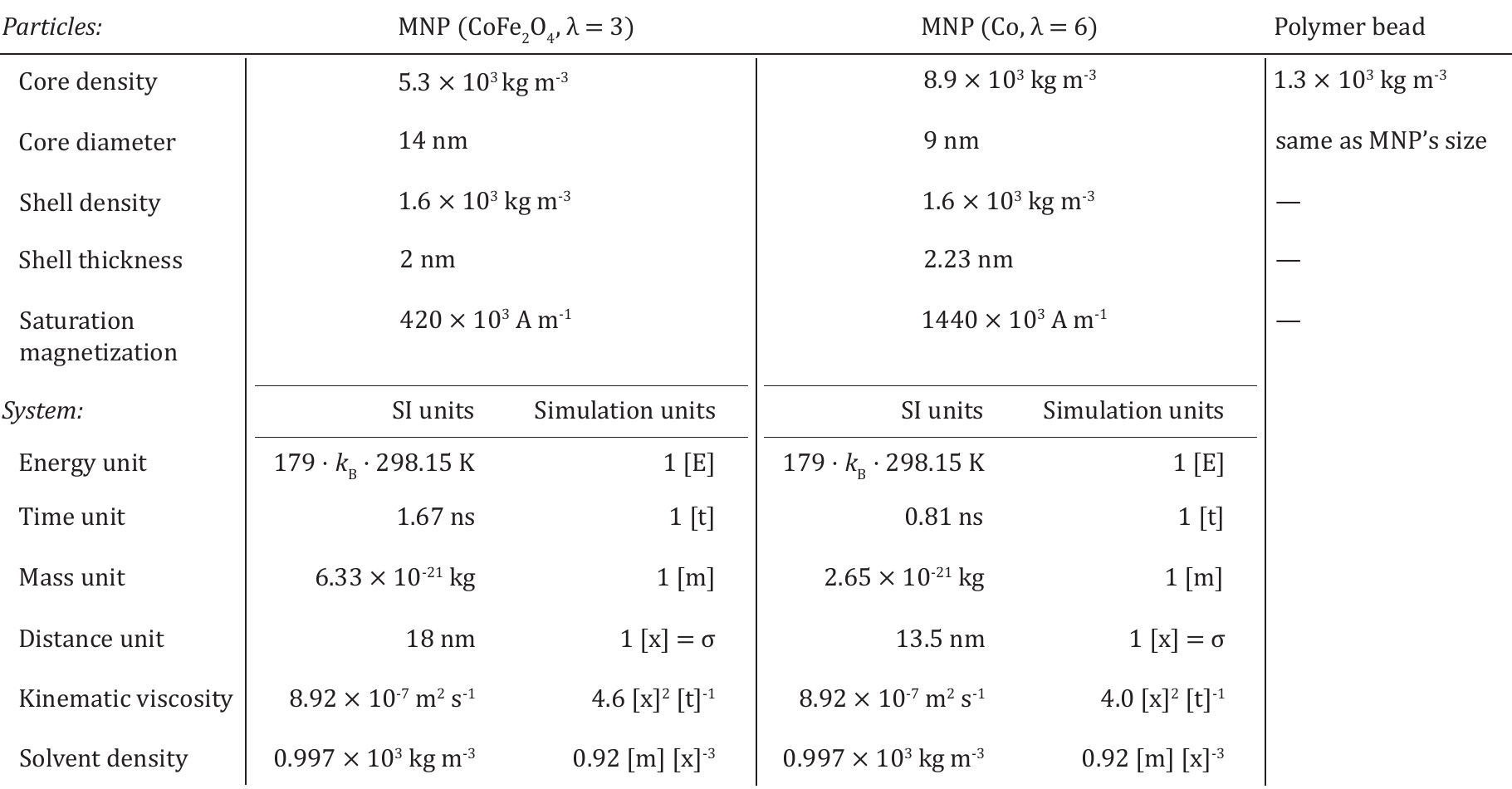}
  \captionof{table}[foo]{Simulation units and their corresponding SI values for two simulated types of gels, which differ by the MNPs contained within as well as polymer beads.}
  \label{fig:table}
\end{figure*}

For convenience, one needs to introduce simulation unit (SU) system, corresponding to the typical nanogels. In simulations we chose magnetic particles to define the length scale.  Thus,  MNPs have dimensionless unit diameter $\sigma=1$. Following it, the sizes of the polymer beads are scaled to the sizes of MNPs. The Table \ref{fig:table} shows the SU and their corresponding SI units for the two simulated systems. 

\subsection{Potentials}

There is a steric repulsion between all kind of particles in a gel. It is given by a shifted and truncated Lennard-Jones (LJ) potential, also known as Weeks-Chandler-Andersen (WCA) potential \cite{1971-weeks}:

\begin{equation}
U_{W C A}(r)=\left\{\begin{array}{ll}{4\left[{(\frac{\sigma}{r}})^{-12}-(\frac{\sigma}{r})^{-6}\right]+1,} & {r \leq 2^{1/6}} \\ {0,} & {r > 2^{1/6}}\end{array}\right. ,
\label{eq:wca}
\end{equation}

\noindent where $r$ is the distance between centers of two interacting beads (measured in $\sigma$). LJ potential well depth set to unity (1 [E]).

The MNG is build out of six independent chains, with $L=100$ beads each. Neighboring beads within the same polymer chain are interconnected by means of finitely extensible nonlinear elastic (FENE) springs tied to their centers, creating the polymer backbone. FENE potential has the form:
\begin{equation}
U_{F E N E}(r)=-\frac{1}{2} \epsilon_{f} r_f^2 \ln \left[1-\left(\frac{r}{r_{f}}\right)^{2}\right],
\label{eq:fene}
\end{equation}
\noindent with the dimensionless bond rigidity $\epsilon_{f}=10$ and the maximum bond extension $r_{f}=1.5$. The chains are crosslinked in a spherical cavity whose size corresponds to the volume fraction $\phi_p \approx 0.1$. Once the system is equilibrated inside the sphere, we choose beads from different chains and connect their centers with elastic springs \eqref{eq:harm} according to a minimum interparticle distance criteria, up to a concentration of cross-linkers $\phi_{\mathrm{links}}=0.17$. 

\begin{equation}
U_{h}(r)=-\frac{1}{2} Kr^{2}.
\label{eq:harm}
\end{equation}
Here, $K=10$ is a dimensionless elastic constant. The spherical confinement removed, when the crosslinking is accomplished.

After that we select randomly 60 polymer beads and place a point dipole of the fixed length in their centres. The size-range considered in this paper justifies this approach, as both cobalt ferrite and cobalt nanoparticles of respectively 18 nm and 13.5 nm in diameter will relax following the Brownian mechanism, rather than N{\'e}el one \cite{ota19a}. The interaction between two magnetic particles can be, thus, described by the dipole-dipole potential:

\begin{equation}
U_{d d}\left(\vec{r}_{i j}\right)=\frac{\left(\vec{\mu}_{i} \cdot \vec{\mu}_{j}\right)}{r^{3}}-\frac{3\left(\vec{\mu}_{i} \cdot \vec{r}_{i j}\right)\left(\vec{\mu}_{j} \cdot \vec{r}_{i j}\right)}{r^{5}},
\label{eq:dipdip}
\end{equation}

\noindent where $\vec \mu_i$, $\vec \mu_j$ are the respective dipole moments of particles that interact and $\vec r_{ij}$ is the vector connecting their centers. The standard way to characterize the interaction between two MNPs is to use the dipolar coupling parameter $\lambda$, which is the ratio of the maximum dipole-dipole interaction energy at contact to the thermal energy, what can be defined as $\lambda=\mu^2$. In this study we have selected MNPs in such a way that $\lambda=3$ (cobalt ferrite) or $\lambda=6$ (cobalt), and further we use $\lambda$ as a reference to a gel with a certain type of MNP inside. 

\subsection{Simulation Protocol}
In this section we discuss the main tools and details of the simulation methods that we employed to investigate MNGs.

Our main tool is the molecular dynamics \cite{2002-frenkel}. As explained above, the MNG is coarse-grained. In order to investigate the effects of the shear flow on such systems, one needs to have an approach for modelling the solvent. Among possible techniques \cite{lusebrink16a,cerda19a,2019-rovigatti-sm-rev,weiss2019spatial,formanek21a}, we opt for the Lattice-Boltzmann scheme \cite{pagonabarraga2004lattice,2008-dunweg}. This method simulates the flow of a Newtonian fluid by solving the discrete Boltzmann equation on a lattice, which corresponds to the Navier-Stokes equations in the limit of small Mach numbers. In this approach, the background fluid is represented by virtual ``effective'' particles that are distributed in space on the nodes of a regular lattice, taking the form of discrete velocity distribution functions. The fluid evolves by applying the discrete Boltzmann equation to the velocity distribution functions. The lattice discretisation scheme underlying the LB method makes it computationally very efficient and easily paralellisable, particularly for graphics processing units (GPU) implementations. For the coupling of MNGs to the fluid, we use the  scheme of Ahlrichs and D\"unweg \cite{2008-dunweg}. In this approach the MNG is thermalised by the fluid. It is worth mentioning, however, that each polymer bead as well as each magnetic particle are point particles, meaning that their rotations are not seen by the fluid. While being a reasonable approximation for polymers \cite{kreissl2021frequency}, for magnetic particles it is a crucial limitation. We overcome it by using Brownian dynamics for thermalisation of dipoles rotation. Magnetic interactions in this setup are calculated directly. We profit from using the simulation package ESPResSo \cite{Weik2019ESPResSoSystems}. 

\begin{figure}[h!]
	\centering
	\includegraphics[width=0.33\textwidth]{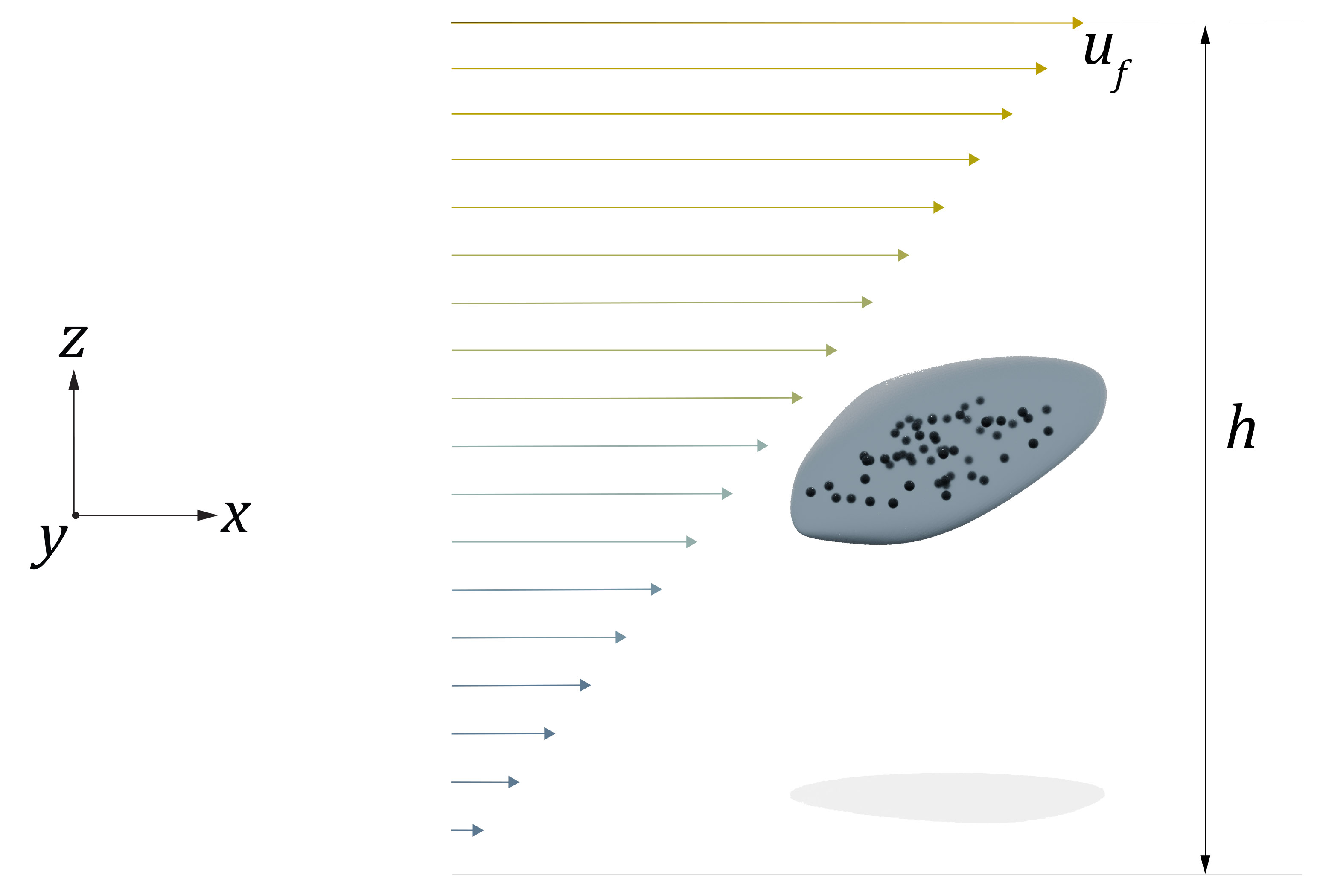}
	\caption{Geometry of the system: the upper wall of a slit with the height $h$ moves with the velocity $\mathbf{u}_f$. Equilibrium configuration of the gel is placed inside and let evolve.}
	\label{fig:sys-geom}
\end{figure}

As shown in Fig. \ref{fig:sys-geom}, we consider a MNG in a slit with two walls separated by distance $h$ along $z$-axis. The upper wall of the box is moving along $x$-axis, creating a linear shear fluid velocity profile, with the shear rate $\dot{\gamma}$. The MNG in its equilibrium configuration is put in the middle of the channel with flow and let to evolve in time until the stationary state is established. 

We use a cube-shaped simulation box with a side length of 100$\sigma$ and the LB lattice constant is set to $a_{grid} = 1 \sigma$.
The system propagates with a fixed time step $\delta t=0.01$, updating LB fluid field at every MD step. When analysing the data, the first  $\SIrange{1}{2.5}{\times 10^{4}}$  simulation steps (depending on shear rate) are not considered. This period of time corresponds to the first oscillation of the gel under the influence of the shear. The total duration of the simulation ranged from $1\times10^{5}$ steps for the fastest shear rate, to $2.5\times10^{5}$ steps for the slowest shear. All measurements are averaged over ten different MNG configurations with distinct magnetic particle positions and cross-linkers distributions, in order to avoid dependence on the individual MNG topology.

The range of shear rates was chosen in such a way that Reynolds numbers 
\begin{equation}
Re_{\gamma}=\frac{|\dot{\gamma}|r^{2}}{\nu},
\label{eq:re}
\end{equation}
\noindent with $\nu$ being the kinematic viscosity and $r$ is the characteristic size of the colloid, are of the order of $10^{-2}$, assuming that $r$ equals to the gyration radius of a gel. The Weisenberg number 
\begin{equation}
Wi=\dot{\gamma}\tau_{r},
\label{eq:wi}
\end{equation}
\noindent for our system lies between $0.1$ and $3$. Here, $\tau_{r}$ is a relaxation time of MNG in no flow condition. The calculation of $\tau_{r}$ is described in the next section. 

In the summary of this section we would like to underline that even in equilibrium without an applied magnetic field, the presence of MNPs alters the structural properties of the nanogels, so we expect the self-assembly of MNPs also manifest itself in the flow. Note that our model is applicable for MNGs below one micron in size, and, for equilibrium, it captures  correctly the formation of MNP bridges. Moreover, we use the crosslinker concentration typical to existing MNGs \cite{Backes2015}. All this gives us the ground to use it for the study of the dynamics below.

\section{Results and Discussions}
\subsection{The influence of shear on the shape of the MNG} \label{sec:shape-eval}

\begin{figure}[h!]
	\centering
	\subfigure[]{\label{fig:traj}\includegraphics[width=0.25\textwidth]{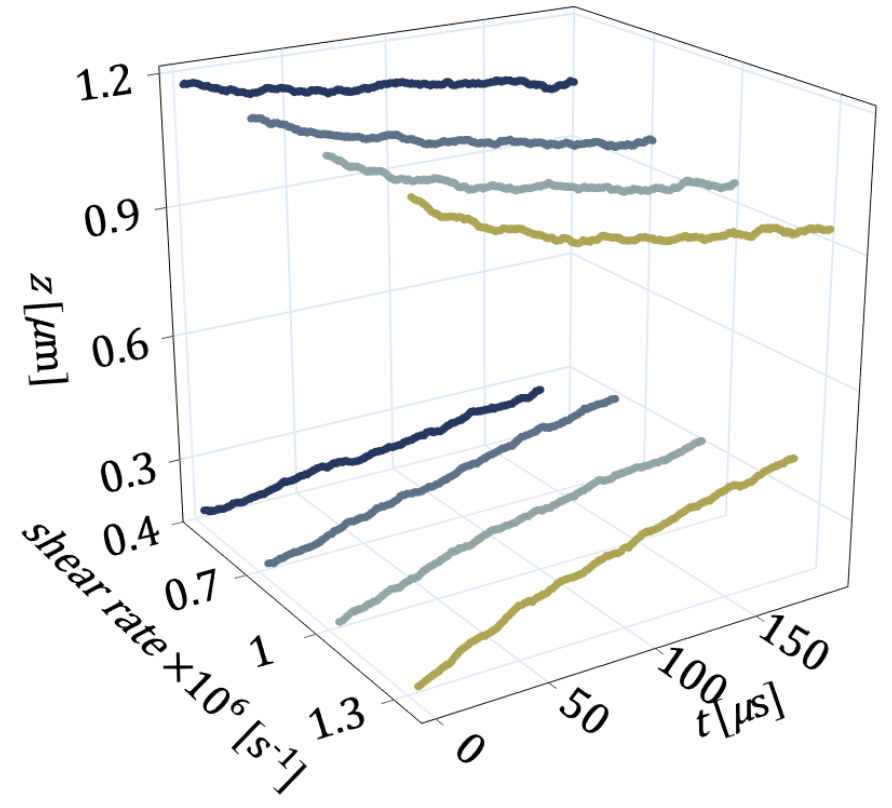}}
	\subfigure[]{\label{fig:Q_vs_h}\includegraphics[width=0.22\textwidth]{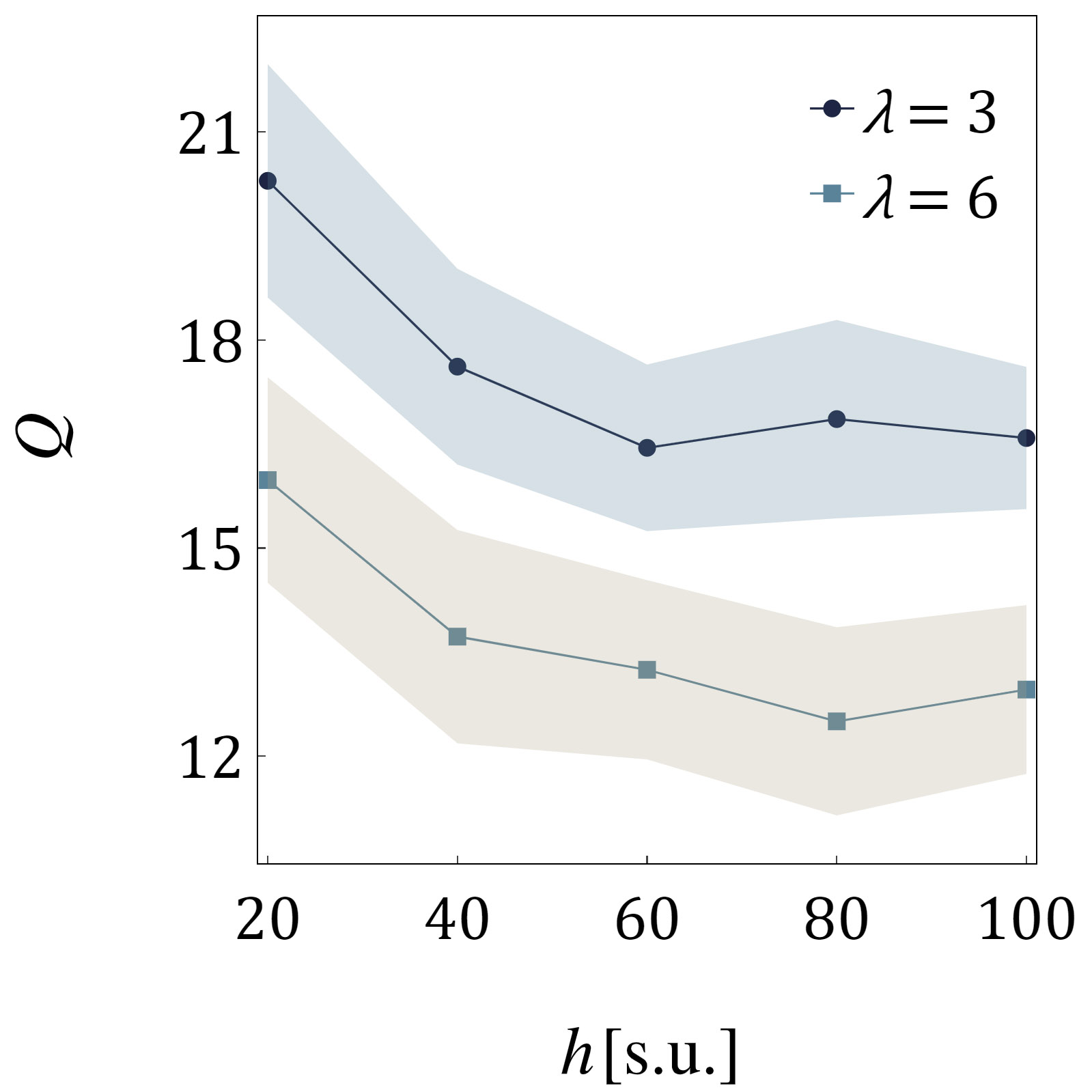}}
	\caption{(a) The trajectory of the MNG centre of mass projected on the $z$-axis for eight independent simulations are presented  as functions of time $t$, $\lambda =3$. MNG starts near the top and near the bottom walls under four different shear rates as shown along the left horizontal axis. (b) Asphericity $Q$ {\it versus} channel size for two different values of $\lambda$ as shown in the legend (effect of the WLF). The halos around the curves show the standard deviation.}
	\label{fig:syst-height}
\end{figure}

Firstly, we analysed the position of the MNG in the stationary state. In Fig. \ref{fig:traj}, the results for eight independent simulations are presented. In each of the realisation, the MNG is initially put at the top or at the bottom of the channel and its centre of mass is followed in time. One can see that for any shear rate indicated by colour and shown along abscissa and for any initial position indicated along the vertical axis the centre of mass of the MNG with time (ordinate) moves to the centre of the channel. Once reaching the stationary position the MNG centre of mass is only fluctuating along $z$-direction.

The stationary position of the MNG in our system, in which no gravity is present, is defined by the wall lift force (WLF). The latter is a result of the increase of the density of flow-lines once the object approaches the wall (for details, see \hyperref[sec:app-A]{Appendix A}). Thus, the stationary state of the MNG is equally distant from both channel walls. The WLF not only defines the stationary position of the MNG, but also might affect its shape, as shown in Fig. \ref{fig:Q_vs_h}, where we plot the asphericity $Q$

\begin{equation}
Q=q_{1}-\frac{1}{2}\left(q_{2}+q_{3}\right),
\label{eq:Q}
\end{equation}

\noindent  with $q_i$ being the eigen values of the MNG gyration tensor, of an MNG in zero flow as a function of channel height $h$ for two different values of the interparticle magnetic interaction $\lambda$. The error-bars are shown with the colour halo around each curve.  One can see, that for any channel with $h<60$ the gel is compressed, while for broader channels the influence of the wall levels out.

\begin{figure}[h]
\centering
  \includegraphics[width=1\columnwidth]{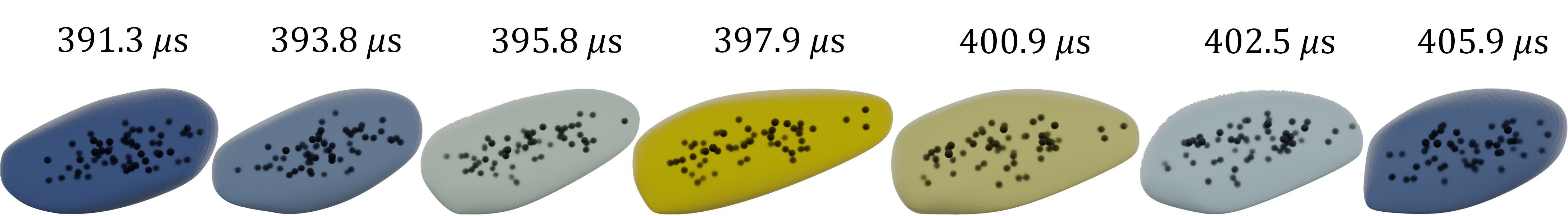}
  \caption{Wobbling of the MNG; $\lambda = 3$ at $\dot{\gamma} = 0.89 \times10^{6} [s^{-1}]$. The colour gradient indicates the degree of asphericity. The corresponding time instances are shown above each configuration.}
  \label{fig:wobbling}
\end{figure}

Looking at the trajectories (see Supplementary video) one notices the oscillatory motion of the gel. The time-line of snapshots is provided in Fig. \ref{fig:wobbling} in order to visualise the changes in MNG asphericity indicated by colour. 

\begin{figure}[h!]
	\centering
	\subfigure[]{\label{fig:Q_freq_vs_shear_rate_SI}\includegraphics[width=0.23\textwidth]{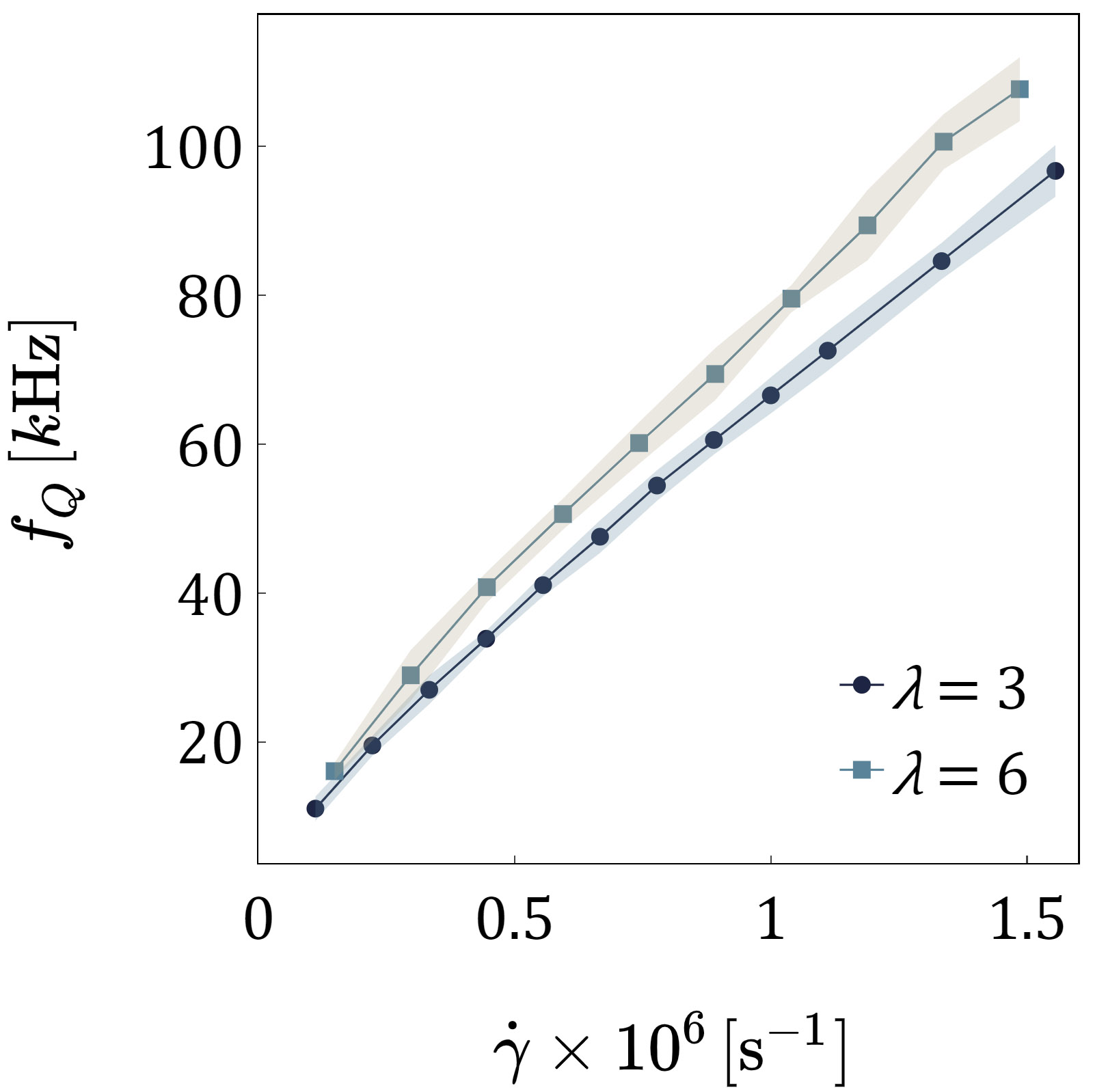}}
	\subfigure[]{\label{fig:Q_vs_shear_rate_SI}\includegraphics[width=0.23\textwidth]{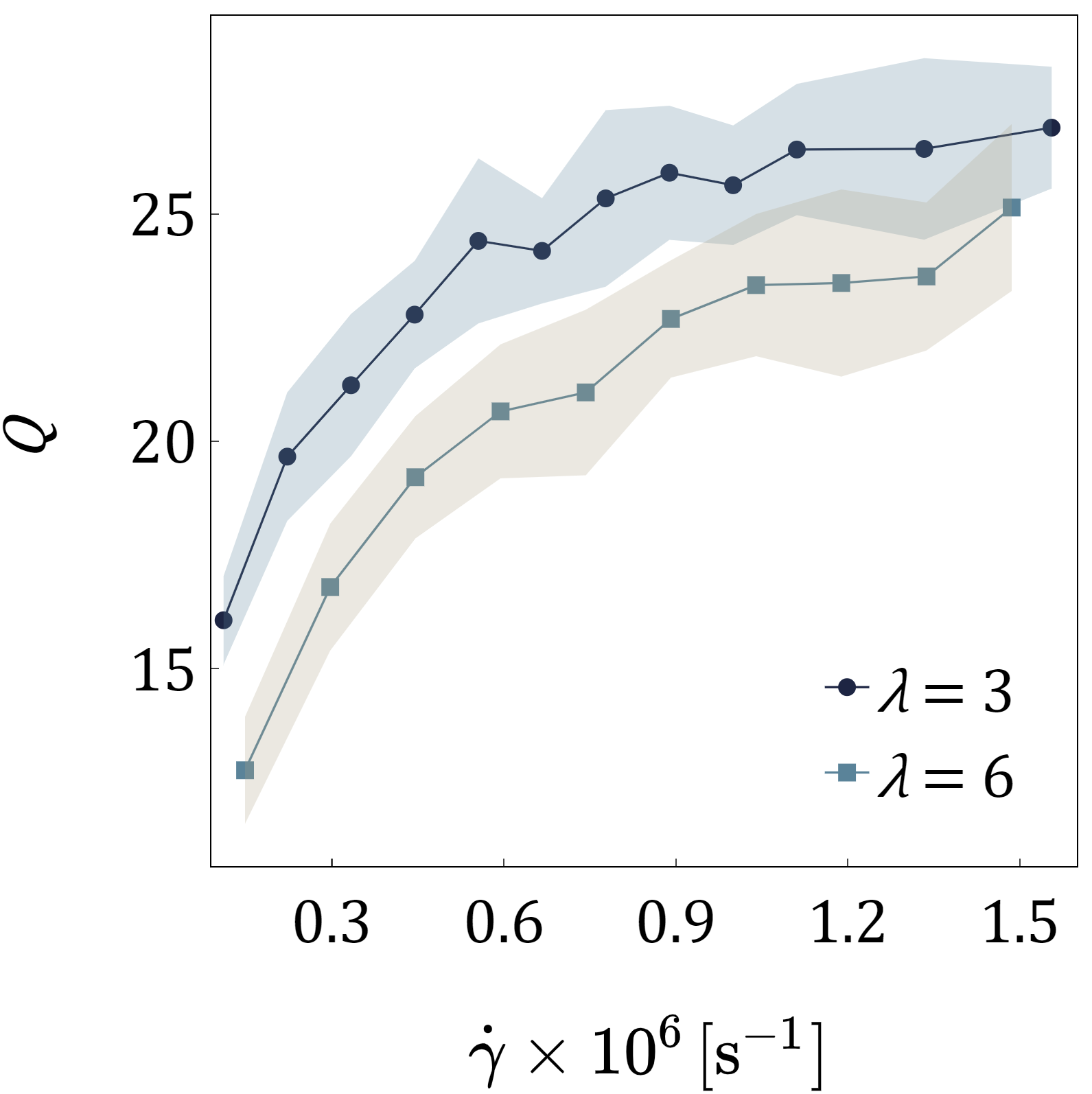}}
	\caption{(a) Frequency $f_Q$ of asphericity oscillations {\it versus} shear rate in SI units. (b) Average value of the asphericity $Q$ {\it versus} shear rate in SI units. Two curves correspond in both plots to different $\lambda$ as shown in the legend. In both plots the errorbars are shown with halos. }
	\label{fig:Q_freq_vs_shear_rate}
\end{figure}

\noindent Here, one can see that first the gel has a mildly elongated shape, then,  with time, MNG elongates and, at the end, it shrinks back as indicated by the change of the colour from dark blue for smaller $Q$ to light khaki for the highest elongation. This motion is periodic and the frequency $f_Q$ of the oscillations is growing with shear rate $\dot{\gamma}$ as shown in Fig. \ref{fig:Q_freq_vs_shear_rate_SI}. Here, we use SI units. It can be seen that the frequency is higher for the MNG with $\lambda =6$. In both cases the frequency is a slowly saturating function of $\dot{\gamma}$.

\subsection{Behaviour of the MNG polymer matrix in the shear flow}\label{sec:matr}

The amplitude of the oscillations, namely the average value of $Q$ saturates faster with growing shear rate as shown by Fig. \ref{fig:Q_vs_shear_rate_SI}. The higher is the magnetic interaction in the MNG the lower is the asphericity. This indicates that there must be some periodic process on the level of the MNG matrix.

\begin{figure}[h!]
	\centering
	\subfigure[]{\label{fig:otlr_vs_time_l3}\includegraphics[width=0.23\textwidth]{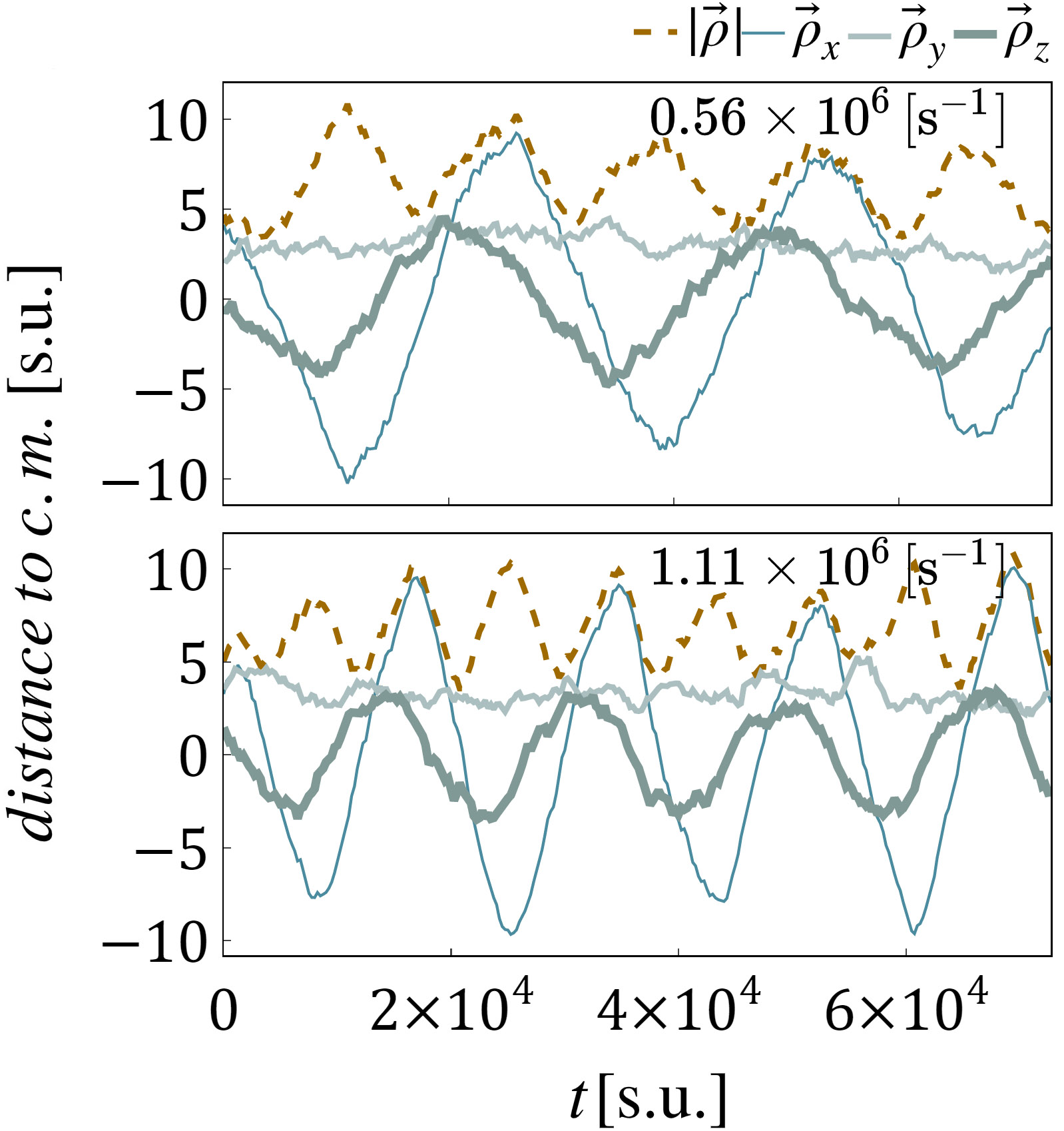}}
	\subfigure[]{\label{fig:otlr_vs_time_l6}\includegraphics[width=0.23\textwidth]{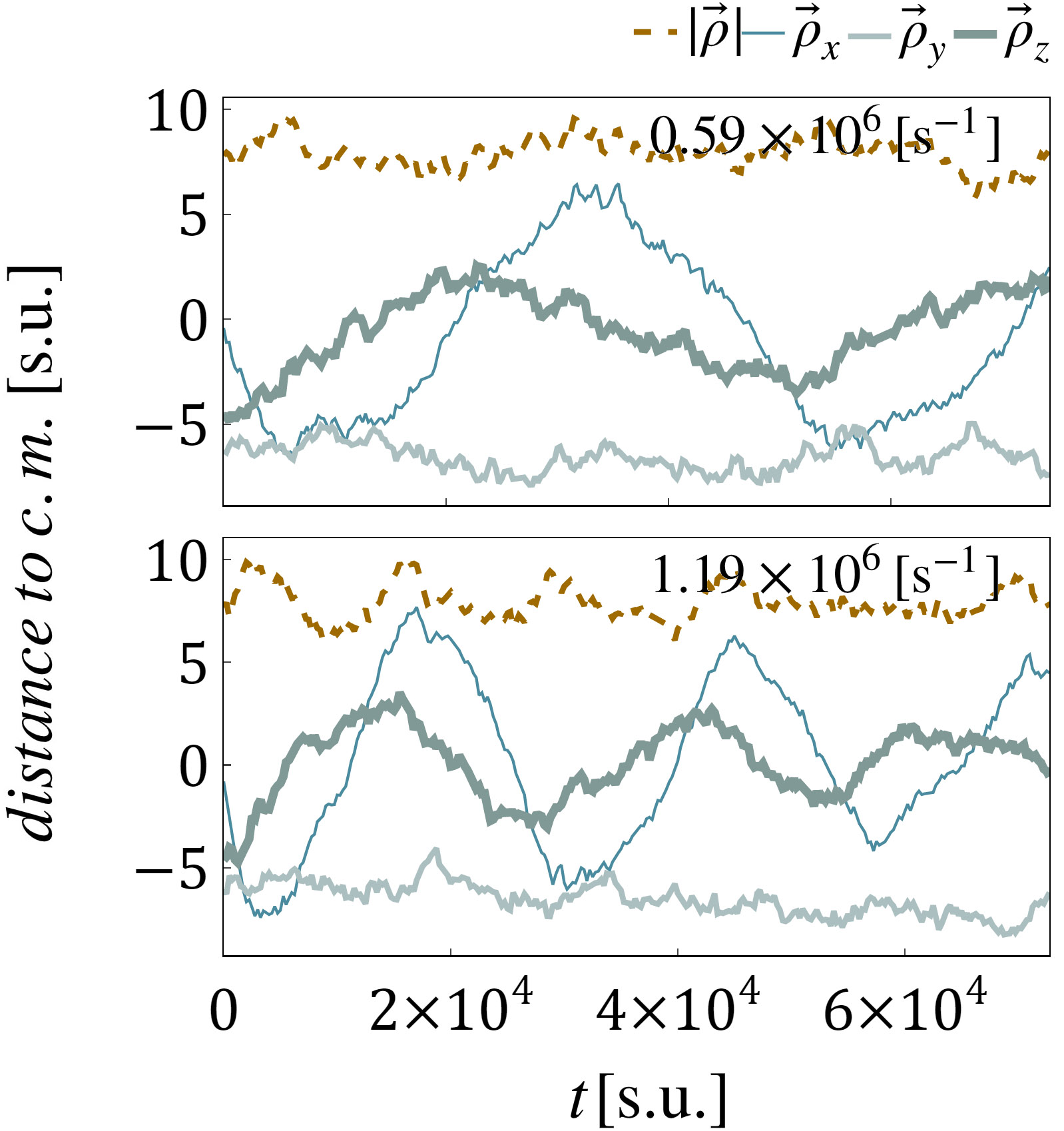}}
	\caption{Time dependence of the modulus of the distance between a randomly chosen particle in the gel and the centre of mass, $|\vec{\rho}|$ (khaki dashed line); of the projections of $\vec{\rho_x}$ on the $x$-axis (highest amplitude, light blue thin line); of the projections of $\vec{\rho_y}$ on the $y$-axis (smallest amplitude, light gay middle thick line); of the projections of $\vec{\rho_z}$ on the $z$-axis (medium amplitude, dark blue thickest line).   (a) $\lambda =3$; top: $\dot{\gamma}=0.56\times 10 ^6$  s$^{-1}$ and below -- 1.11$\times$10$^6$ s$^{-1}$. (b) $\lambda = 6$; top: $\dot{\gamma}=0.59\times 10 ^6$  s$^{-1}$ and below -- 1.19$\times$10$^6$ s$^{-1}$}
	\label{fig:otlr_vs_time}
\end{figure}

Such a process can be analysed by looking at Fig. \ref{fig:otlr_vs_time}. Here, we plot how the modulus of the distance between a randomly chosen particle in the gel and the centre of mass, $|\vec{\rho}|$ (khaki dashed line) and the projections of $\vec{\rho}$ on the three axis (indicated in the legend) change in time. In Fig. \ref{fig:otlr_vs_time_l3} the results are plotted for the MNG with $\lambda= 3$. Here, in the plot above the shear rate is 0.56$\times$10$^6$ s$^{-1}$ and below -- 1.11$\times$10$^6$ s$^{-1}$. In Fig. \ref{fig:otlr_vs_time_l6} analogous curves but for the MNG with cobalt nanoparticles , {\it i.e.} $\lambda = 6$ are presented. In the top plot $\dot{\gamma}$ is almost the same as in the top plot of Fig. \ref{fig:otlr_vs_time_l3}; similarly, in the lower plot the shear rate is twice higher. In both cases, perpendicular to the flow direction only small fluctuations are observed, however, one clearly sees the oscillations in $z$ and $x$ directions. So the particle periodically approaches and goes away from the MNG centre of mass. Importantly, the frequency of these oscillations is both shear rate and $\lambda$ dependent.

\noindent The frequency of the oscillations in Fig. \ref{fig:otlr_vs_time} coincides in all four cases with the corresponding frequencies at which $Q$ is oscillating (Fig. \ref{fig:Q_freq_vs_shear_rate_SI}), also confirmation of these words can be seen further in Fig. \ref{fig:v_vs_shear_l3} and \ref{fig:v_vs_shear_l6}. This leads us to the conclusion that these two motions are two sides of the same tumbling-wobbling motion performed by the MNG around its centre of mass.

\begin{figure}[h!]
	\centering
	\subfigure[]{\label{fig:angle_acfs}\includegraphics[width=0.48\textwidth]{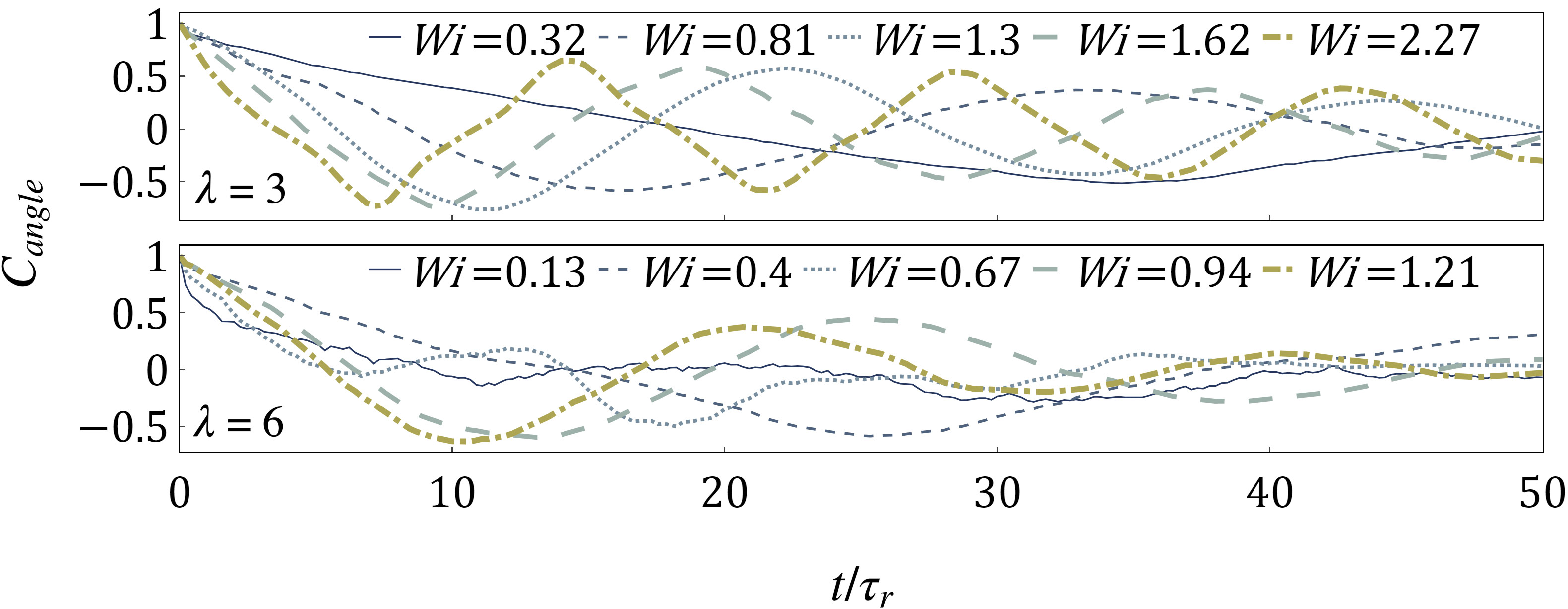}}
	\subfigure[]{\label{fig:freq-to-shear}\includegraphics[width=0.23\textwidth]{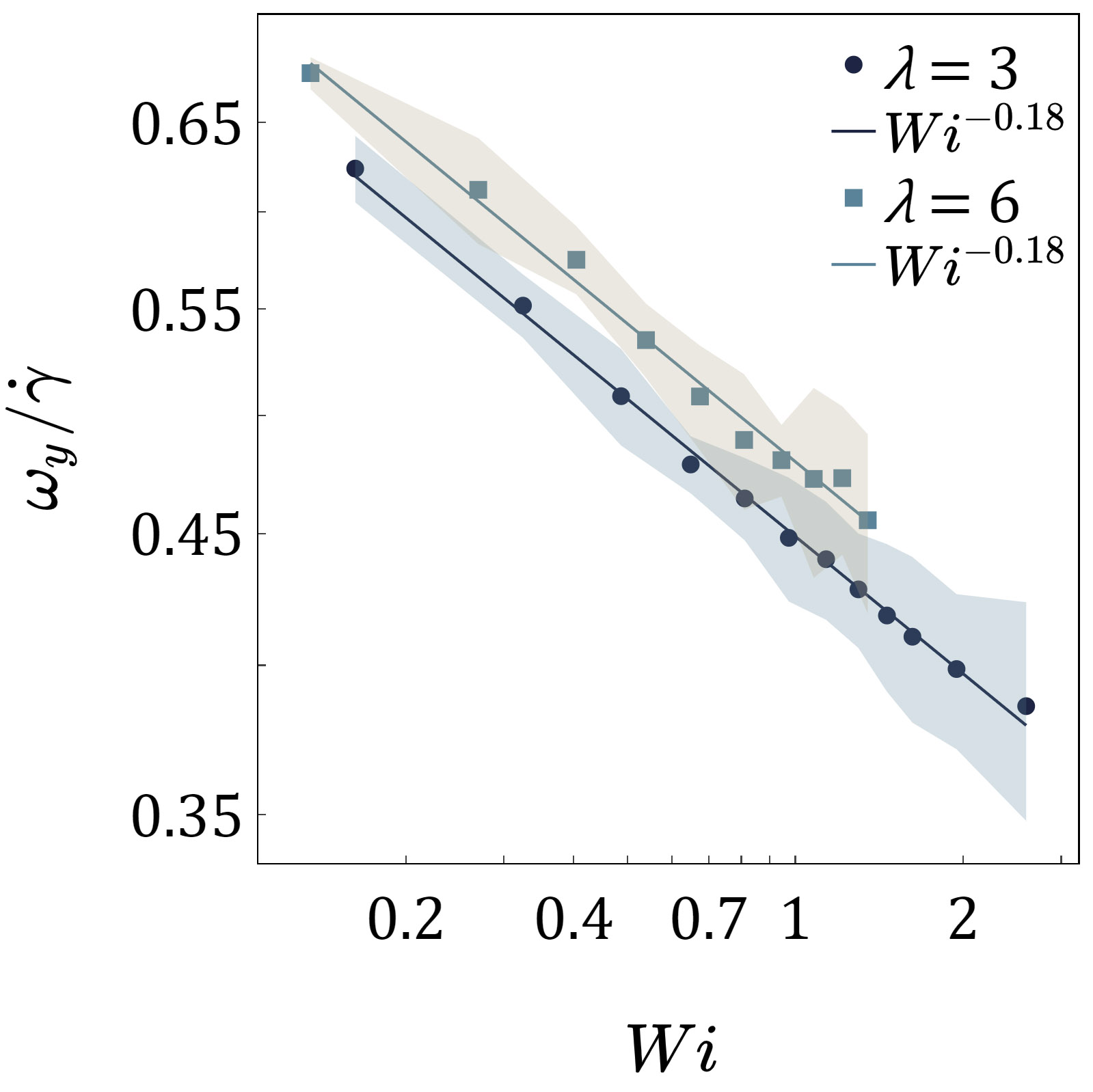}}
	\caption{(a) Angular auto-correlation functions, $C_{angle}$, {\it versus} rescaled time for different $Wi$ as shown in the legend. Top: $\lambda=3$; Bottom: $\lambda =6$. (b) Rescaled angular velocity of the MNG as a function of $Wi$ for different values of $\lambda$ in log-log scale. Solid lines are power law fits, halos show the errobars.}
	\label{fig:angle_acf}
\end{figure}

In order to look deeper into the rotations of a MNG, in Fig. \ref{fig:angle_acf}, we plot the orientational auto-correlation functions, $C_{angle}$. This function is built for the angle between the eigen vector of the MNG gyration tensor corresponding to the largest eigenvalue, {\it i.e} the MNG anisotropy axis, and vector $\vec{\rho}$ {\it versus} time, normalised by the eigen relaxation time. Different curves correspond to various values of the Weisenberg number. Here, one can see that the correlations decay faster for higher $Wi$, particularly for $\lambda = 6$ shown in Fig. \ref{fig:angle_acfs} bottom plot. The period of the correlations for similar values of $Wi$ is smaller for MNG with cobalt ferrite. As the orientational auto-correlations decay faster for the MNG with $\lambda =6$ cobalt nanoparticles than for a more loose and less spherical MNG with cobalt ferrite MNPs, one can summarise that the MNG with higher magnetic interactions due to its more compact and rigid shape with lower asphericity is less affected by the flow. It is worth saying that if one scales the curves in Fig. \ref{fig:angle_acfs} along the $x$-axis by the corresponding $Wi$ (i.e. by the strain), then for $Wi>0.6$ the trends collapse, showing that once the growth of the average apherisity, $Q$, shown in Fig. \ref{fig:Q_freq_vs_shear_rate}, reaches the plateau, the overall motion regime of the MNG stabilises as well.

Generally, one can conclude that MNGs with cobalt nanoparticles are more rigid than their counterparts with less magnetic particles resulting in smaller deformations and faster rotations. Another important observation is that the rotation and deformation of the MNG occur with the same frequency. 

Importantly, even the MNGs with cobalt particles investigated here remain very soft and highly deformable. In Fig. \ref{fig:freq-to-shear}, we show the ratio of the rotational frequency to the shear rate as a function of $Wi$. As expected, both for $\lambda = 3$ and $\lambda =6$, the curves exhibit a power-law decay with increasing $Wi$. For a sphere the limit of small Weisenberg numbers is one half \cite{snijkers09a}. The absolute values corresponding to the MNGs are much smaller and exhibit no initial plateau. On the one hand, Fig. \ref{fig:freq-to-shear} shows how strongly the MNGs are affected by the flow. On the other hand, looking at the fitting, we understand that the difference in MNPs interactions does not affect the scaling of the plotted decay, making the part of MNPs in the MNG less clear and worth understanding in more detail. So, in the next section, the influence of the shear rate on the magnetic particles inside the MNGs is discussed.

\subsection{The competition between magnetic and hydrodynamic interactions}\label{sec:magn}
\begin{figure}[h!]
	\centering
\subfigure[]{\label{fig:dip_vs_time_l3}\includegraphics[width=0.23\textwidth]{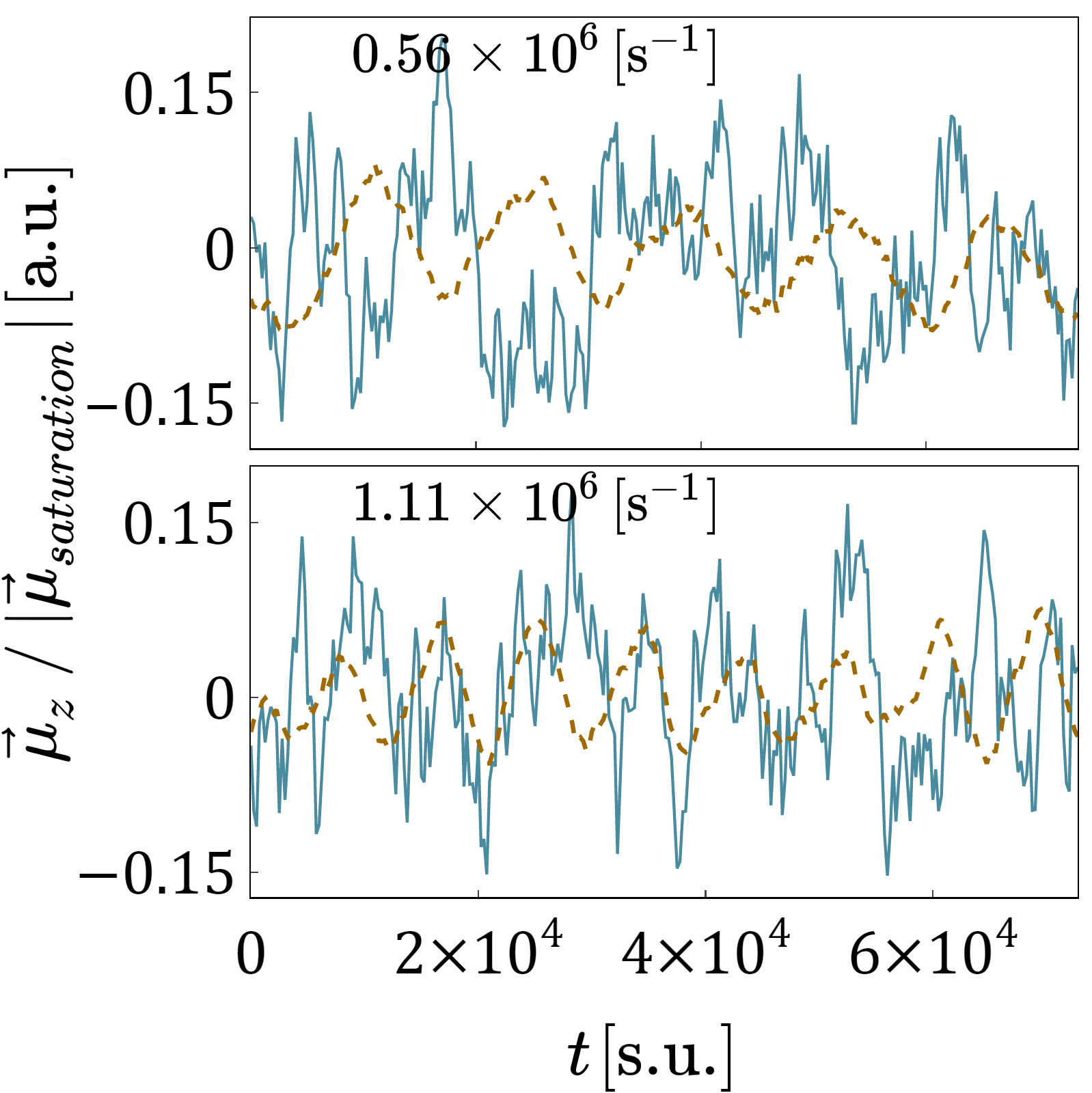}}
\subfigure[]{\label{fig:dip_vs_time_l6}\includegraphics[width=0.23\textwidth]{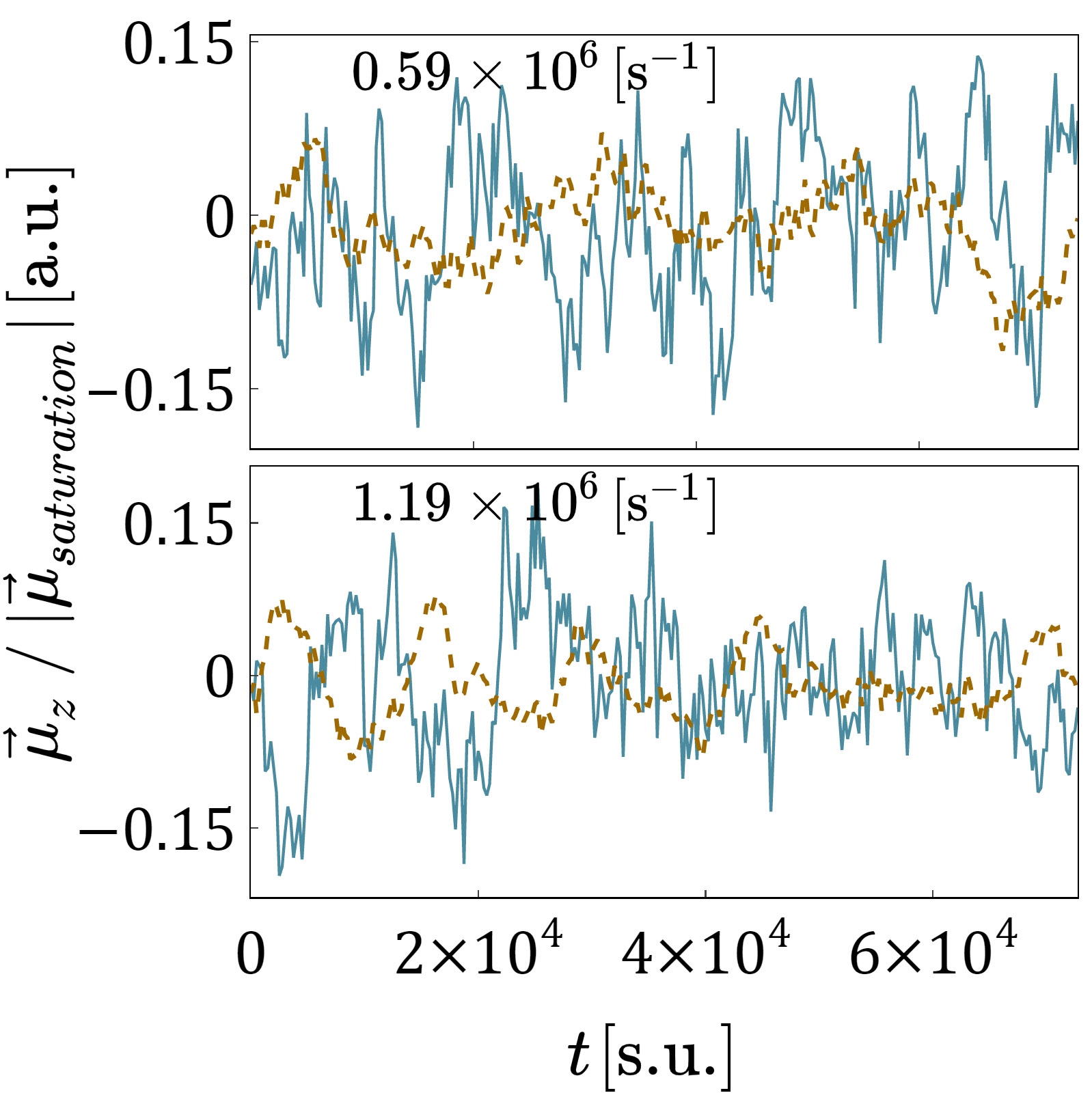}}
\subfigure[]{\label{fig:dip_vs_shear}\includegraphics[width=0.23\textwidth]{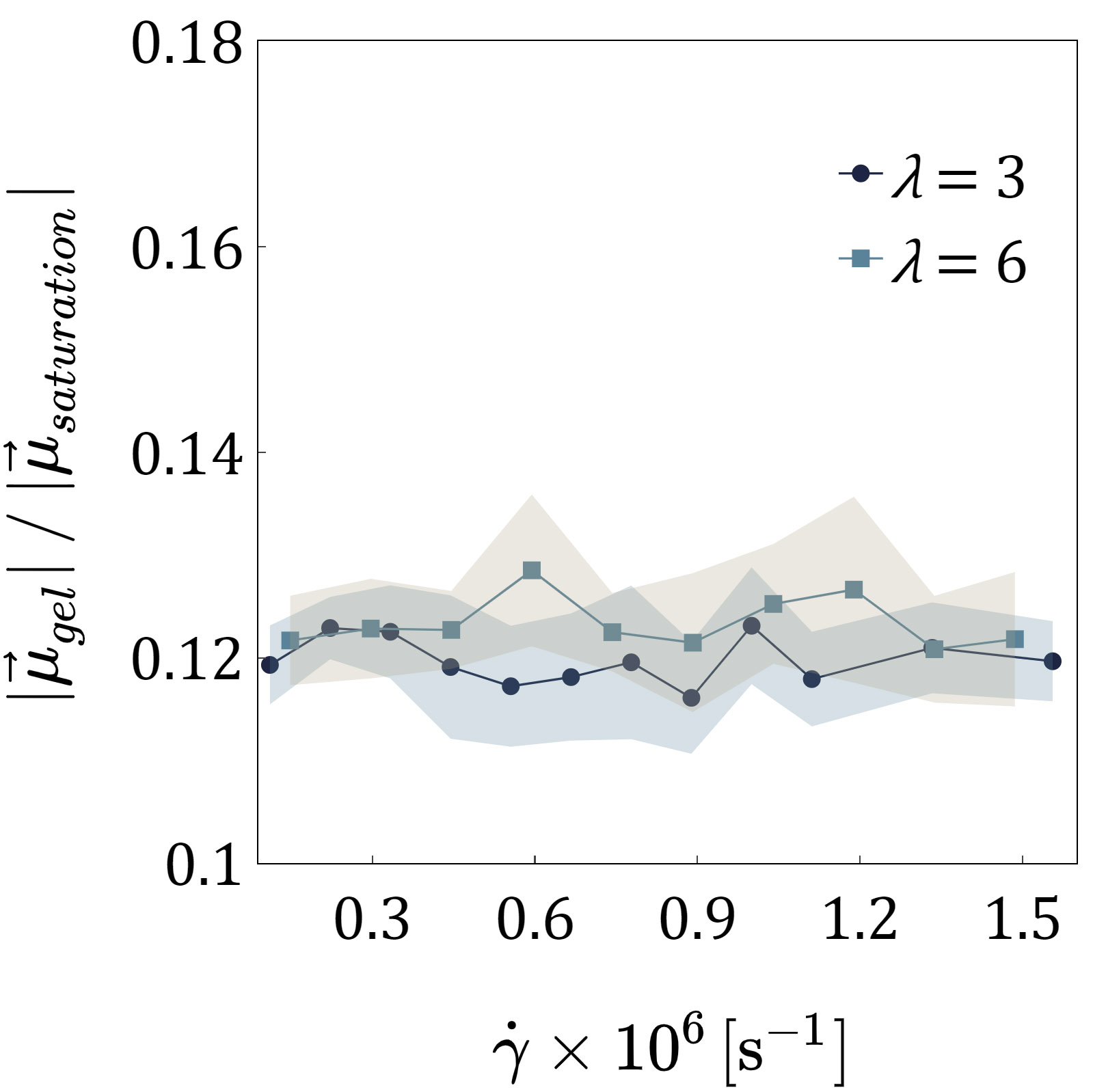}}

	\caption{ (a)-(b)Time dependence of the $z$-axis projection of the MNG total magnetic moment $\vec{\mu}_{z}/|\vec{\mu}_{saturation}|$, blue. (a) $\lambda =3$; top: $\dot{\gamma}=0.56\times 10 ^6$  s$^{-1}$ and below -- 1.11$\times$10$^6$ s$^{-1}$. Khaki dashed lines here show rescaled and shifted oscillations of $|\vec{\rho}|$:  $a(|\vec{\rho}|-\langle|\vec{\rho}| \rangle) +b$, the mean coincides with the mean of the blue curve and the amplitude is half of that. They serve to compare characteristic frequencies of the oscillations.  (b) $\lambda = 6$; top: $\dot{\gamma}=0.59\times 10 ^6$  s$^{-1}$ and below -- 1.19$\times$10$^6$ s$^{-1}$. (c) The modulus of the total MNG magnetisation normalised by the corresponding saturation value {\it versus} shear rate.}
	\label{fig:dips}
\end{figure}
First of all we check whether the  magnetic moment of the MNG follows the tumbling-wobbling motion and also oscillates in time. As it can be seen in Figs. \ref{fig:dip_vs_time_l3} and \ref{fig:dip_vs_time_l6}, where the projection of the total magnetisation on the $z$-axis normalised by the saturation magnetisation of the MNG, $\vec{\mu}_{z}/|\vec{\mu}_{saturation}|$, is plotted as a function of time for different shear, it is not fully the case. The khaki dashed line is the rescaled and shifted $|\vec{\rho}|$ from Figs. \ref{fig:otlr_vs_time_l3} and  \ref{fig:otlr_vs_time_l6} respectively. Those curves serve to compare the magnetisation oscillations to that of the MNG shape. One can see that only for $\lambda =3$ the projection $\vec{\mu}_{z}/|\vec{\mu}_{saturation}|$ oscillates in time with the same period as $|\vec{\rho}|$, albeit rather noisily. As for the case of $\lambda =6$, both oscillations seam to be much less pronounced.   Moreover, from Fig. \ref{fig:dip_vs_shear}, one can conclude that  independently from the magnetic interactions the normalised reminent magnetisation of the MNG,$|\vec{\mu}_{gel}|/|\vec{\mu}_{saturation}|$, is rather small and barely changes with the shear rate.

\begin{figure}[h!]
\centering
\subfigure[]{\label{fig:clusters-hist-l3}\includegraphics[width=0.23\textwidth]{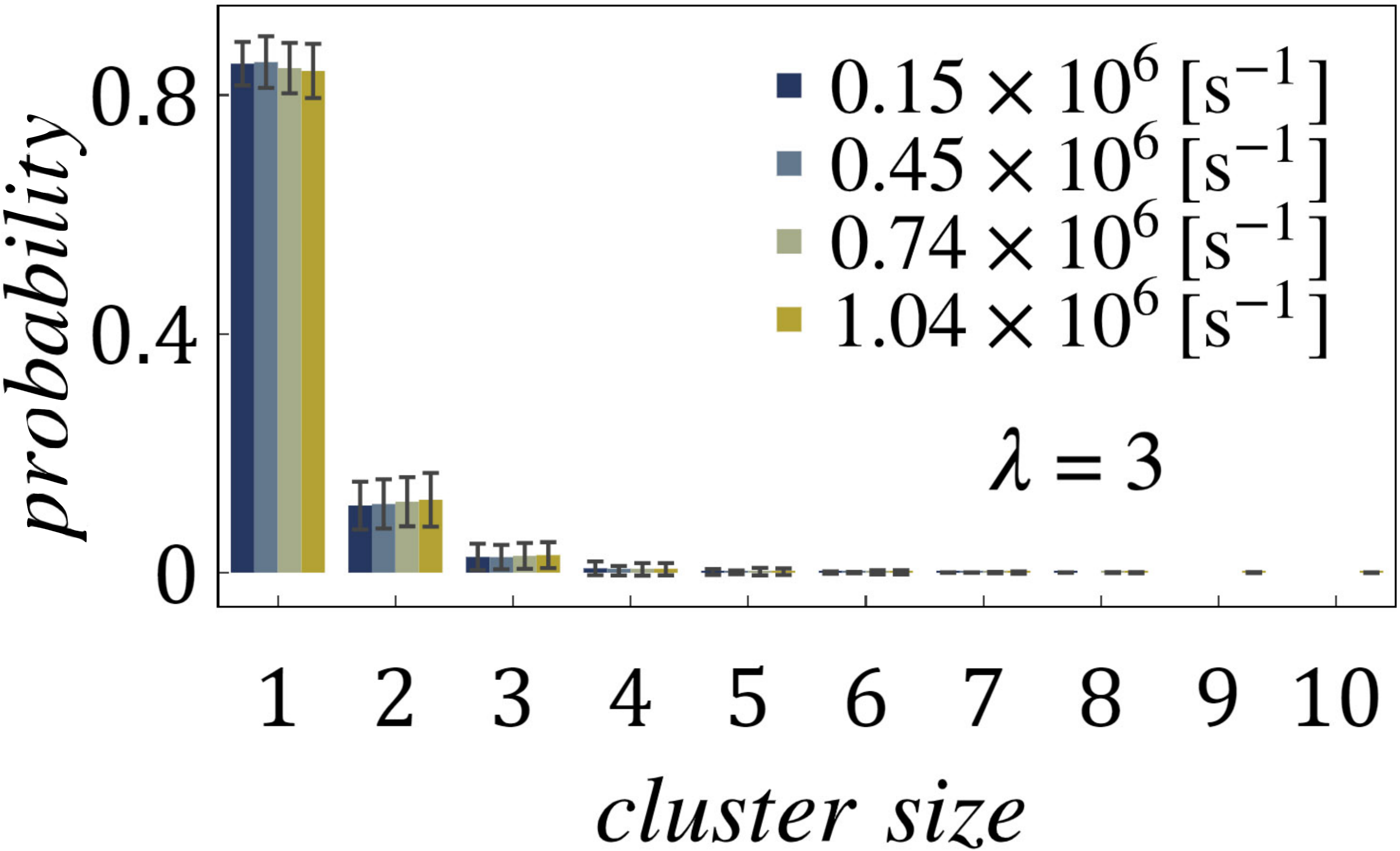}}\\
\subfigure[]{\label{fig:clusters-hist-l6}\includegraphics[width=0.23\textwidth]{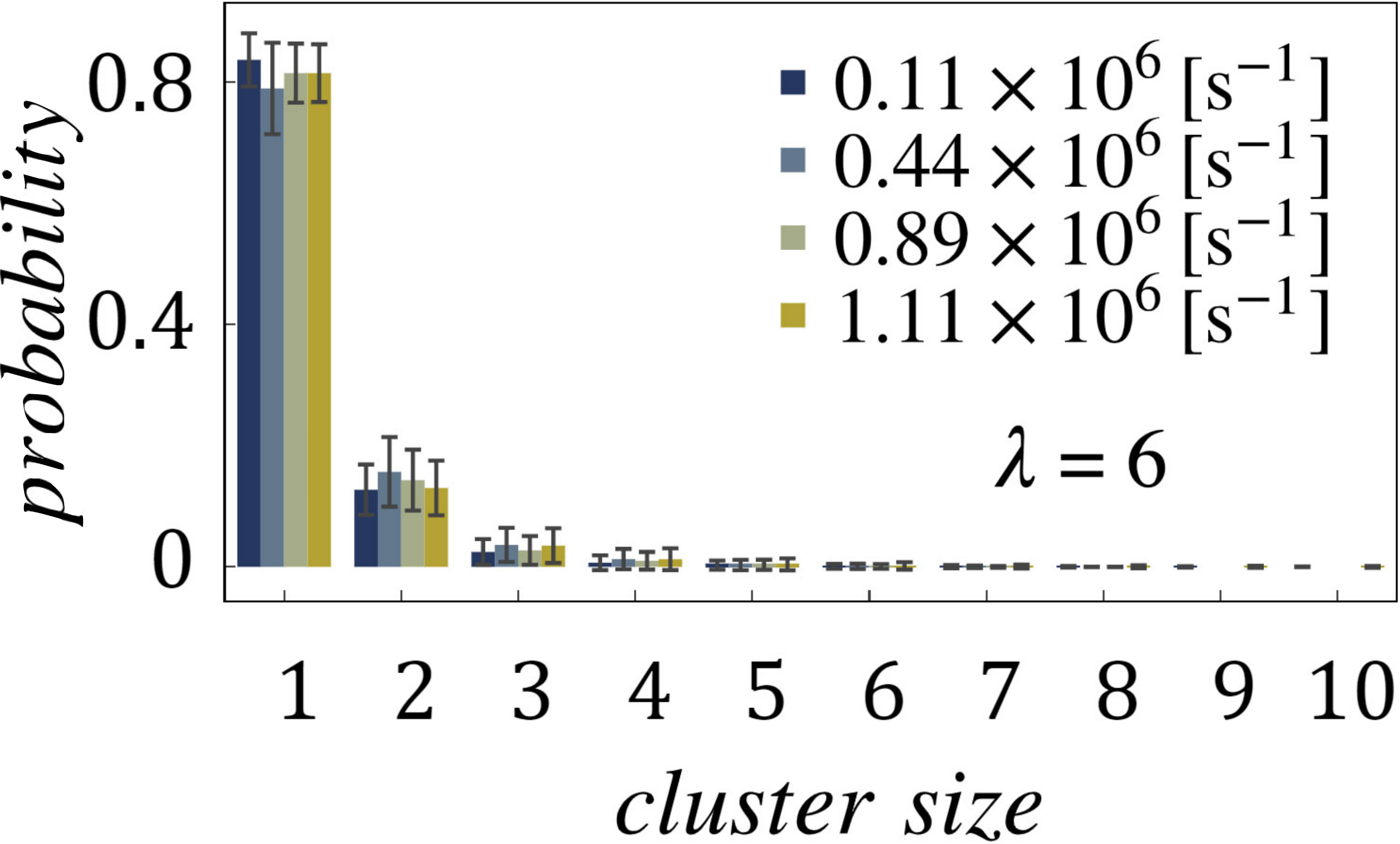}}
	\caption{Histograms, showing the probability to find a MNP cluster of a given size for different values of $\dot{\gamma}$, as shown in the legend. (a): $\lambda = 3$. (b) $\lambda =6$. }
	\label{fig:clusters-hist}
\end{figure}

In zero shear, magnetic particles in MNGs, particularly in case of $\lambda = 6$,  tend to form clusters, optimising this way the dipolar energy.  Even though the cluster size is lower than that in bulk suspensions due to elastic constraints, cobalt nanoparticles at the studied concentration form chains whose length is two-four particles on average. Had the clusters persisted in the flow and rotated together with the polymer matrix one could have expected the total magnetic moment to have a more sound periodic behaviour, particularly for high values of $\lambda$.  As the histograms of cluster sizes, shown in Figs. \ref{fig:clusters-hist-l3} and \ref{fig:clusters-hist-l6}, illustrate, independently from the magnetic interaction strength considered in this study, even the smallest hydrodynamic flow considered here leads to clear destruction of most magnetic clusters, as less than 10 per cent of particles are forming dimers, not to mention larger clusters whose probability is below five percent. 

\noindent Independently from $\lambda$, the majority of particles remains single. For $\lambda =6$ as shown in Fig. \ref{fig:clusters-hist-l6}, the influence of the shear is rather small, and a very small fraction of relatively large clusters can be found for any flow. The tendency is different for the MNG with $\lambda =3$. Looking at Fig. \ref{fig:clusters-hist-l3}, it can be observed that the higher the shear rate the more dimers can be found in the MNG. This effect is rather weak, but persistent, however its amplitude might be comparable to the statistical error and needs further investigation provided below.

\begin{figure}[h!]
	\centering
	\subfigure[]{\label{fig:v_vs_shear_l3}\includegraphics[width=0.23\textwidth]{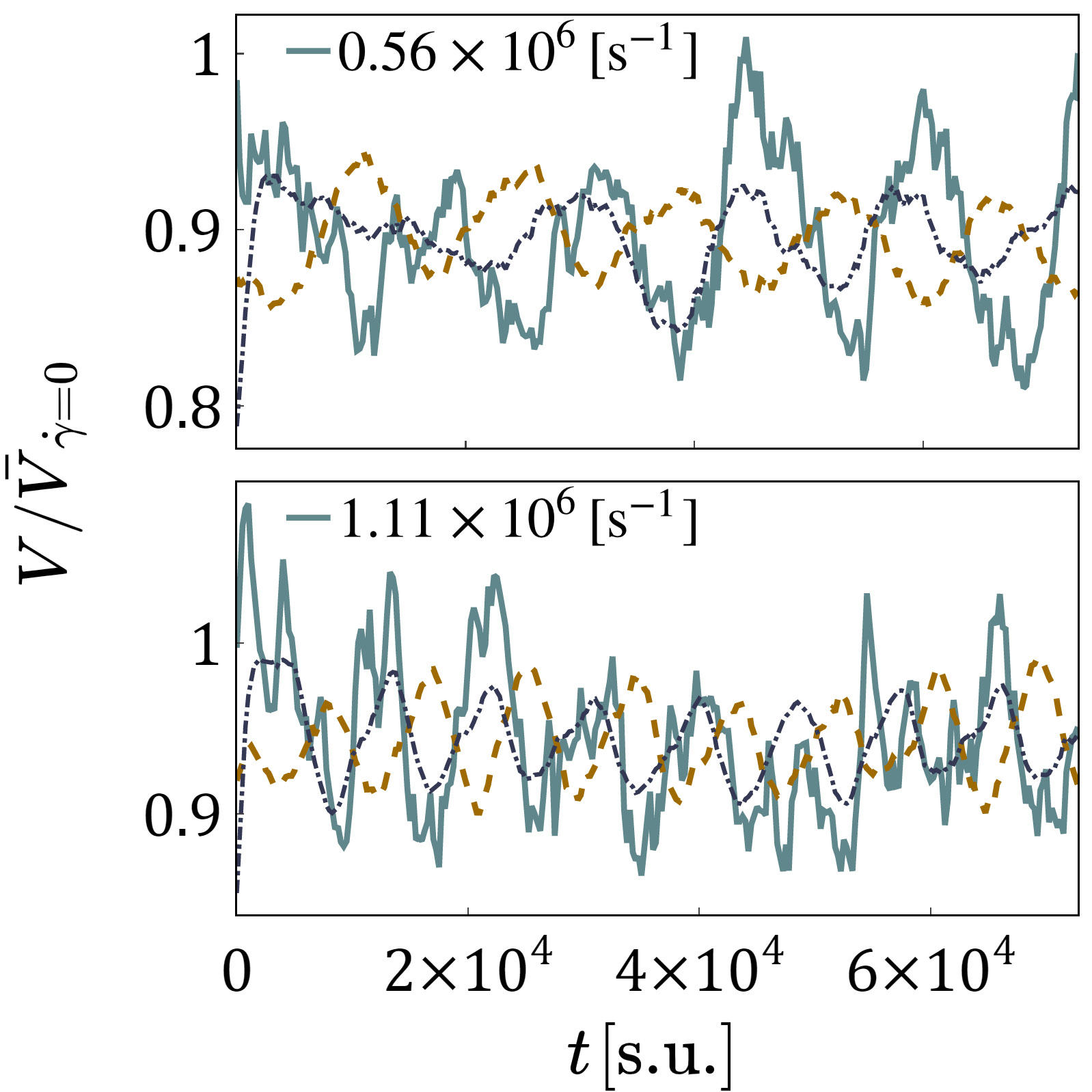}}
	\subfigure[]{\label{fig:v_vs_shear_l6}\includegraphics[width=0.23\textwidth]{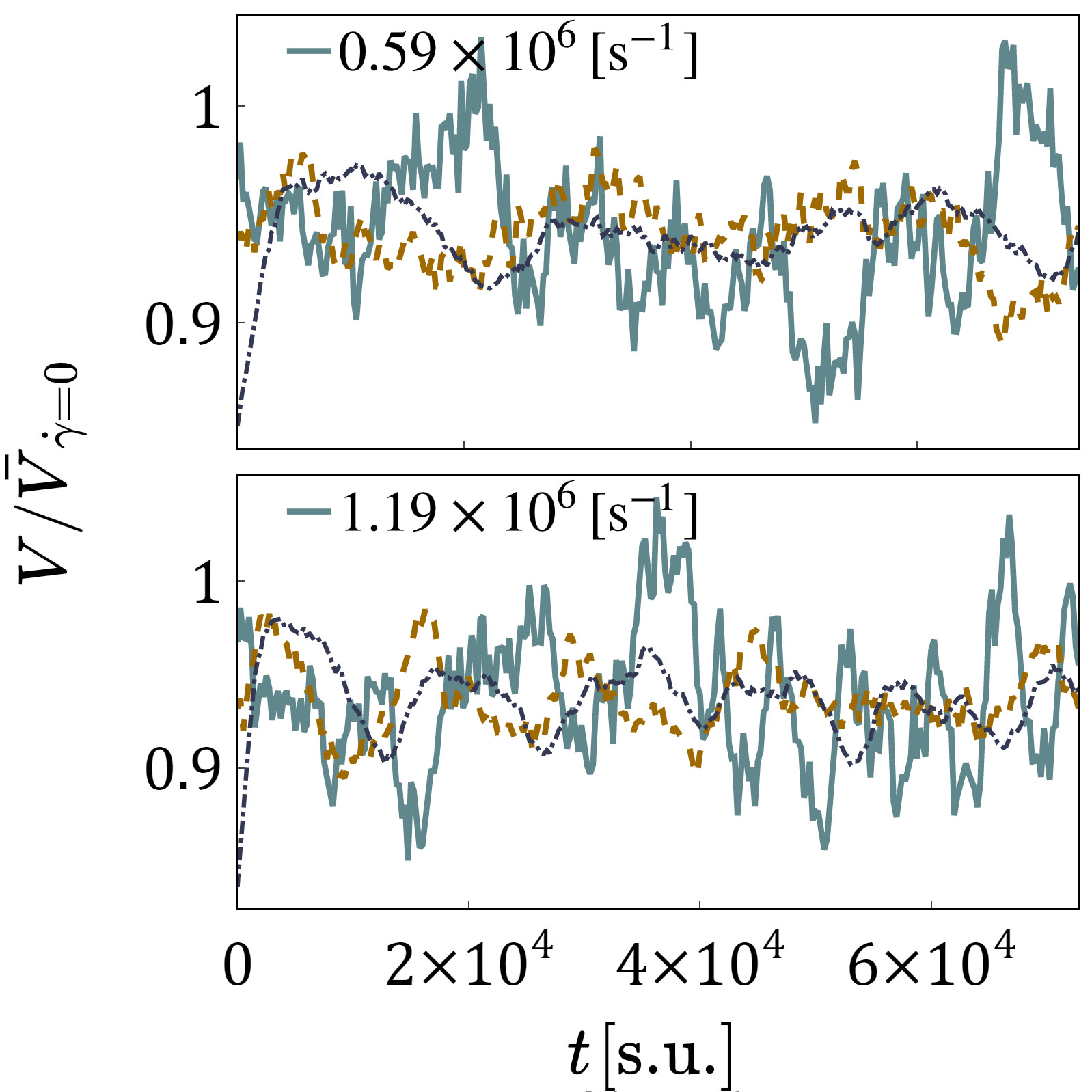}}
	\subfigure[]{\label{fig:v_vs_shear}\includegraphics[width=0.23\textwidth]{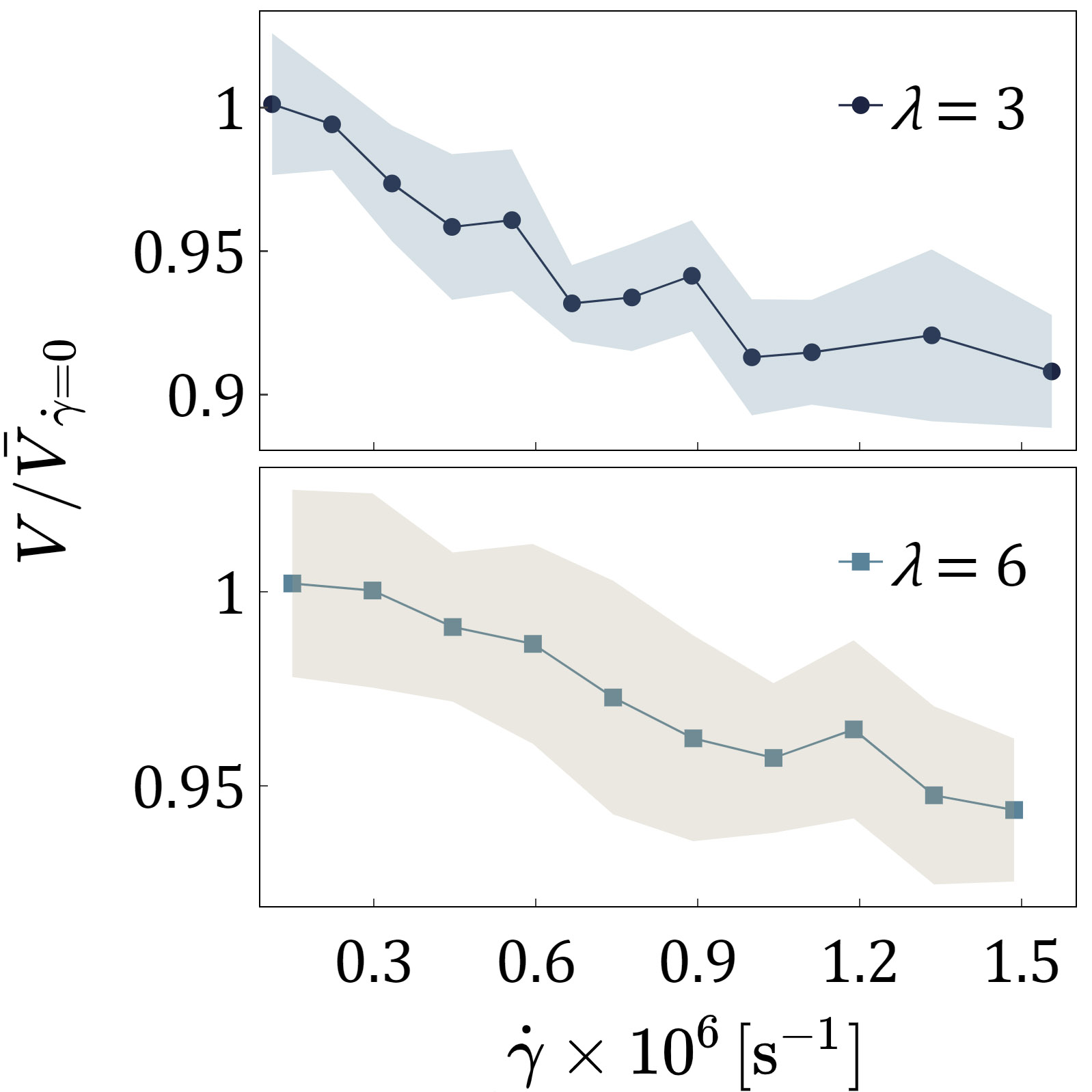}}
	\caption{(a)-(b)Time dependence of the MNG reduced volume $V/\bar{V}_{\dot{\gamma} = 0}$ Khaki and dark blue dashed lines here show oscillations of $|\vec{\rho}|$ and $Q$ correspondingly, both are rescaled and shifted in the same manner, as $|\vec{\rho}|$ in Fig. \ref{fig:dip_vs_time_l3} and \ref{fig:dip_vs_time_l6}. (a) $\lambda =3$; top: $\dot{\gamma}=0.56\times 10 ^6$  s$^{-1}$ and below -- 1.11$\times$10$^6$ s$^{-1}$. (b) $\lambda = 6$; top: $\dot{\gamma}=0.59\times 10 ^6$  s$^{-1}$ and below -- 1.19$\times$10$^6$ s$^{-1}$. (c) Normalised volume {\it versus} shear rate. Top: $\lambda = 3$. Bottom: $\lambda =6$. }
	\label{fig:volume}
\end{figure}

The first factor that usually affects the magnetic nanoparticles self-assembly is the interparticle interaction energy, which in our case is shear flow independent. The second factor that influences MNP self-assembly is the concentration. When investigating magnetic response of the MNGs in thermodynamic equilibrium we found already that due to the higher local concentration of the MNPs in the MNGs their suspension has higher magnetic susceptibility than in case the same volume fraction of MNP is allowed to freely move in the whole volume. At the same time, going back to the section \ref{sec:shape-eval}, we found that the flow changes the shape of the MNG periodically in time and the increase of the shear leads to the increase of the time-averaged asphericity. If the same type of behaviour, namely periodic contraction-expansion in time for a fixed $\dot{\gamma}$ and overall contraction for growing shear rates,  was true for the MNG volume, one could attribute the potential growth of the dimer fraction to the increase of the local MNP concentration. To verify this conjecture, in Fig. \ref{fig:volume} we describe the behaviour of the normalised MNG volume $V/\bar{V}_{\dot{\gamma} = 0}$, where the normalisation factor is the volume of the MNG without the flow. 

\begin{figure}[h!]
	\centering
	\subfigure[]{\label{fig:freq_V_vs_freq_U}\includegraphics[width=0.23\textwidth]{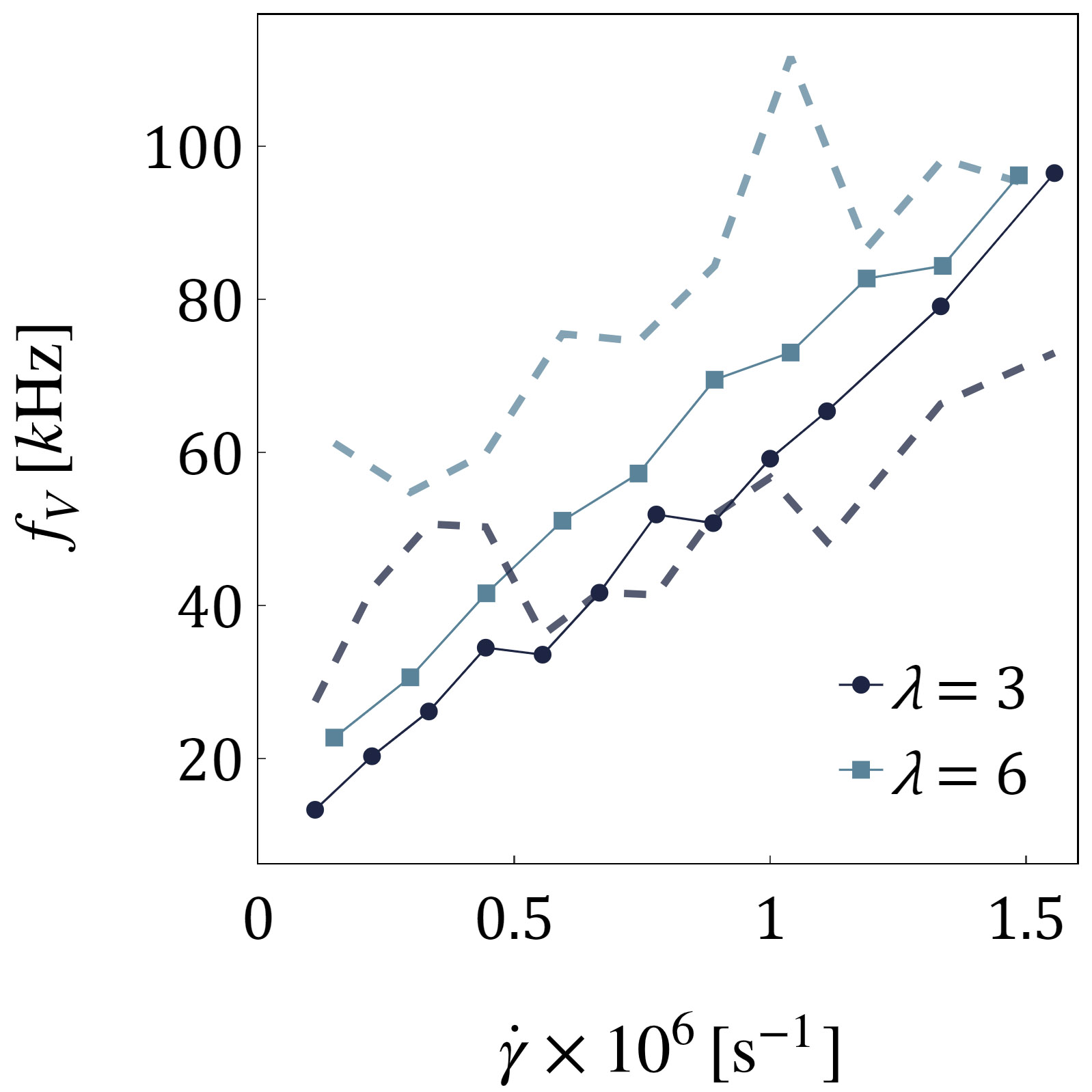}}
	\subfigure[]{\label{fig:dimers}\includegraphics[width=0.23\textwidth]{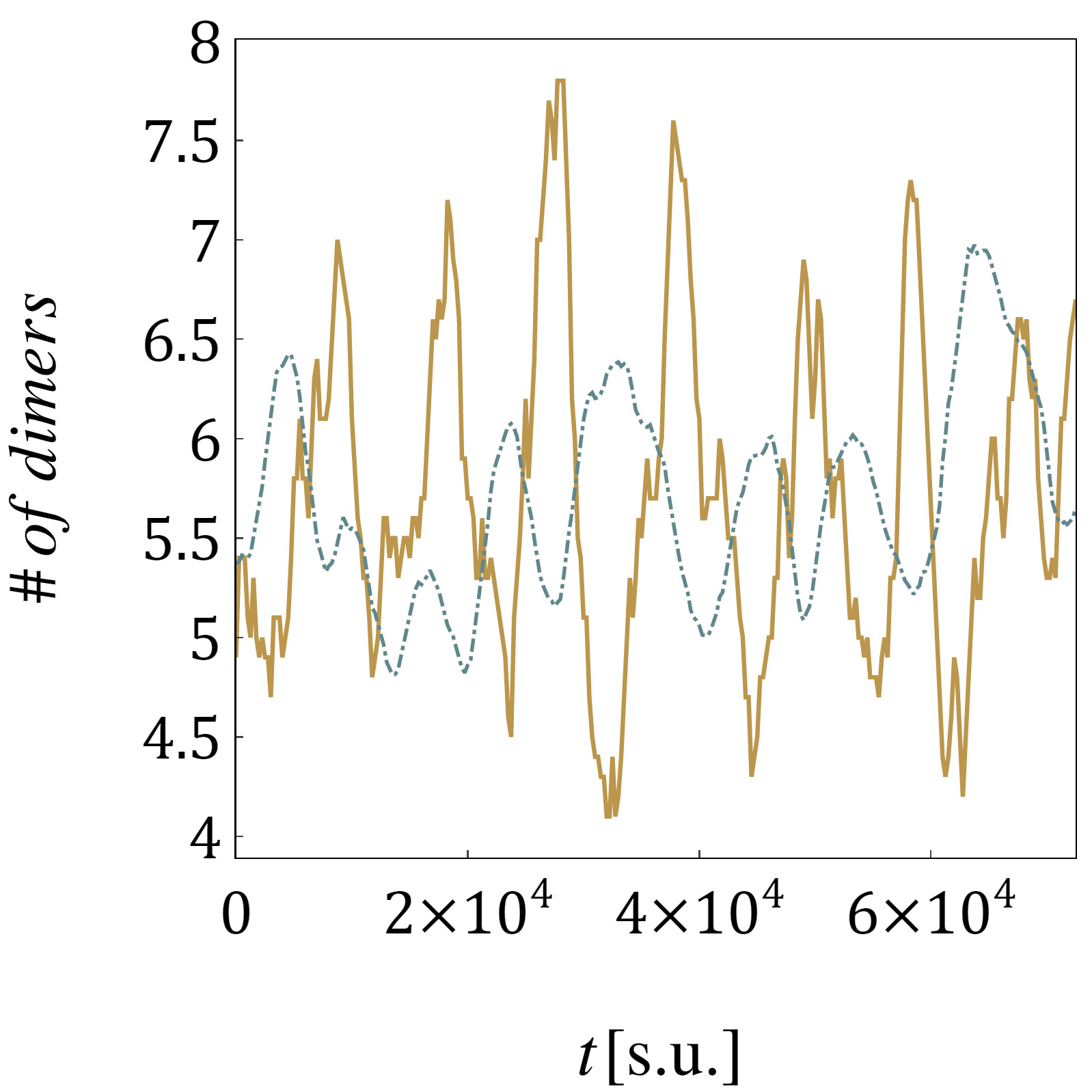}}
	\caption{(a) Frequency $f_V$ of volume oscillations {\it versus} shear rate in SI units. The dashed line corresponds to the frequency of the total energy of the dipole-dipole interaction. (b) Time dependence of number of dimers from MNPs inside MNG; smoothed out by a moving average. Dashed line is for rescaled and shifted volume, in the same way, as $|\vec{\rho}|$ in Fig. \ref{fig:dip_vs_time_l3} and \ref{fig:dip_vs_time_l6}. $\lambda =3$; $\dot{\gamma}=1.49\times 10 ^6$  s$^{-1}$}
	\label{fig:volume_vs_dip}
\end{figure}

Here, in Figs. \ref{fig:v_vs_shear_l3} and \ref{fig:v_vs_shear_l6}, we plot the time evolution of the ratio between the volume of the MNG at a given shear rate to the volume, the MNG has if no flow is present. The volume of the MNG is calculated by triangulating a mesh representation of the MNG and summing up the volumes of the obtained tetrahedrons.  For the case of MNG with cobalt ferrite MNPs, as shown in Fig. \ref{fig:v_vs_shear_l3}, for both high and low shear rates the volume of the MNG oscillates around the value that is slightly below its $\bar{V}_{\dot{\gamma}=0}$. The frequency of oscillations is the same as that of the wobbling shown in Fig. \ref{fig:otlr_vs_time} and coincides with the oscillations of $\mu_z$ shown in Fig. \ref{fig:dip_vs_time_l3}. For higher values of $\lambda$, the oscillations are less well-defined as seen in Fig. \ref{fig:v_vs_shear_l6}, however still present. Fig. \ref{fig:v_vs_shear} evidences that the overall contraction that the MNG with cobalt MNPs exhibits for a given shear rate is smaller than that of it counterpart with cobalt ferrite. Considering the fact that the volume in the latter goes down by approximately ten per cent, the same growth in the local concentration of MNPs inside is to be expected, what could lead to the increase in the number of dimers formed in the MNG and shown in Fig. \ref{fig:clusters-hist-l3}. Even though, as mentioned before, from the first glance, this effect seems to be small and possible to be attributed to the statistical error, it is not the case. In fact, if there is the change in the number of clusters it should be reflect in the total energy of the dipole-dipole interaction ($\sum{U_{dd}}$) (see Eq. \eqref{eq:dipdip}).  More specifically, we are interested in the relationship between the dipolar energy and  the MNG volume. Firstly, in Fig. \ref{fig:freq_V_vs_freq_U}, we present the oscillation frequencies of both quantities depending on $\dot{\gamma}$ (volumes - solid lines with markers; $\sum{U_{dd}}$ - dashed lines). Light blue curves correspond to $\lambda =6$, and dark blue to $\lambda =3$. For both values of $\lambda$, the correlation coefficients between volume and energy: $r_{V, \sum{U_{dd}}} \approx 0.85$, which indicates a clear dependence of those two oscillatory processes. Secondly, if we look at Fig. \ref{fig:dimers}, where number of dimers (solid brown line) inside MNG versus time presented, as well as normalized and shifted MNG's volume (dashed light-blue line), we can observe antiphase oscillations. Thus, even small fluctuations in volume $(<10\%)$ affect the correlations between magnetic nanoparticles and in some cases result in the increase of probability to find a dimer of MNPs previously discussed. More calculations related to the correlation between MNPs volume fraction changes and their self-assembly can be found in  \hyperref[sec:app-B]{Appendix B}.

The same chain of arguments can be constructed for the MNGs with $\lambda = 6$. In fact, the overall contraction is less pronounced, the magnetic interactions are higher, so the shear effects are more screened in these MNGs. 
\subsection{Does MNG in a carrier liquid behave as a droplet in the emulsion?}\label{sec:positioning}
In the 1930s Taylor proposed a successful model describing the droplet deformation in a shear flow \cite{taylor1934formation}. The deformation, obtained from two-dimensional images, was characterised by Taylor's parameter:
\begin{equation}
D=\frac{L-B}{L+B},
\label{eq:D}
\end{equation}
where $L$ is the length of the droplet, and $B$ is its breadth. This expression is very similar to the definition of asphericity in Eq. \eqref{eq:Q} \footnote{If we consider the droplet/gel in the $xz$-plane, then $L$ is nothing but $q_{1}$, and $B$ is $q_{3}$.}. Fig.~12 in the work of Taylor \cite{taylor1934formation} plots the dependence of $D$ on the capillary number, a dimensionless quantity proportional to the shear rate. We have obtained qualitatively the same trend in Fig. \ref{fig:Q_vs_shear_rate_SI}. However, a visual similarity should not be misleading. First of all, a MNG has no surface tension and the carrier liquid can easily go in and out of the polymer mash. Secondly, one has to keep in mind the role of magnetic nanoparticles that affect the rigidity and the tumbling-wobbling frequency of the motion. Finally, a droplet is usually considered to be formed by a Newtonian liquid contrary to non-Newtonian nanogel in our case.

The latter restriction has been partially eliminated in the recent study, in which the deformation of the liquid filled capsules with elastic membranes under shear flow was investigated \cite{Sui2008}. The authors used Lattice-Boltzmann method combined with the immersed boundary concept. In the latter study, the capsule, similarly to our MNG, undergoes periodical shape deformation during its tank-treading motion, however, in Ref. \cite{Sui2008}, it has not been shown that the those motions are just the two sides of the same oscillatory process, as it is clearly the case for MNGs considered here. 

To conclude, it is worth mentioning that the change of the MNG volume with increasing shear rate that plays a crucial part in the magnetic energy oscillations in the present study, and is reaching 10 per cent in its value, has been only briefly mentioned previously as a weak and unessential effect \cite{C7SM01829K, bhattacharjee18a}. 

In other words, despite seemingly standard behaviour, the MNGs investigated here have distinct features not inherent to emulsion droplets, capsules or nonmagnetic MGs.

\section{Conclusions}\label{sec:con}

Using coupled Molecular Dynamics -- Lattice-Boltzmann approach, we investigated the behaviour of a magnetic nanogels in a shear flow. This behaviour, particularly the effects of the flow on the gel shape and internal structure, as well as the ways to fine-tune them, is of particular relevance in microfluidics, as the drug delivery capacity of MNGs directly depends on those factors. 

Note that in this work, we focus exclusively on the behaviour of MNGs in the absence of an externally applied magnetic field. This, on one hand, allows us to focus on the interplay between hydrodynamic and intrinsic magnetic interactions and analyse the eigen deformations of the MNG. On the other hand, field-free transport is of interest for the applications where the flow is present and no magnetic drag is required, such as blood vessels in the direct vicinity of the target. In the latter case the field will be mainly used for keeping the magnetic particles in the target and/or for drug release.

In order to model realistic MNGs, we chose the parameters close to those experimentally available systems \cite{Witt2019}. 

Our investigations have shown that in a shear flow with moderate rates, {\it i.e.} with Wisenberg number on the order of one, the centre of mass of a MNG tends to be in the centre of a channel, even though the flow profile inside is not symmetric. The reason for this is a wall lift force that acts on the gel. This force is weaker than in case of a compact sphere, but still important for a MNG with cross-linker concentration of 17 per cent. It is rather striking, considering that the shape of such MNGs is very flexible and the liquid easily penetrates the body of the gel. 

The flexibility of the shape manifests itself in a combined wobbling-rotation motion that a MNG performs with respect to its centre of mass. Each polymer bead as well as magnetic particle in the MNG rotates around the centre of mass and oscillates with respect to the latter. It results in tumbling and wobbling of the whole MNG. The frequency of the two processes is identical. It grows with increasing shear rate and magnetic interactions.

In order to elucidate the interplay between magnetic and hydrodynamic interactions, MNGs with either cobalt or cobalt ferrite particles under similar flow conditions are investigated. Even though we find that stronger magnetic interactions lead to the growth of the overall rigidity of the gel that results in weaker deformations and faster oscillations, the flows considered in this work largely impede the formation of long magnetic nanoparticle chains that are present in static conditions. We show that due to the fact that the MNG volume oscillates in time, for the MNG with cobalt ferrite particles the total magnetisation oscillates as well. The overall decrease of the volume with growing shear results in the growth of local MNP concentration inside the gel that manifests itself in higher correlations between MNPs.

The next step would be to investigate the influence of an applied magnetic field and verify whether it can offer mechanisms to compete with hydrodynamic interactions in order to preserve magnetic clusters and change the wobbling-rotating scenario. 

\section*{Acknowledgments}
This research has been supported by the Russian Science Foundation Grant No.19-12-00209. Computer simulations were performed at the Vienna Scientific Cluster (VSC). I.S.N. and S.S.K. are grateful to Vienna Doctoral School Physics, Doctoral College DCAMF and were partially supported by FWF Project SAM P 33748.  The authors thank Pedro S. S{\'a}nchez and Dr. Rudolf Weeber for fruitful discussions and useful recommendations.

\section*{Appendix A: Wall lift force (WLF)}\label{sec:app-A}

Looking at the behaviour of the magnetic nanogel in the vicinity of the wall, we found that the MNG was always pushed away towards the channel centre. As mentioned in Section \ref{sec:shape-eval}, the drift towards the centre can be demonstrated by placing the gel initially near the top or the bottom wall and following its transverse motion perpendicular to the flow direction.

Over the past 50 years, not less than 20 experimental, analytical and computational studies have been devoted to the WLF, see works  \cite{wlf_low,wlf_heigh} and references therein. We also could not resist the temptation to measure this effect in our system. We chose WLF as a benchmark to check the accuracy of the LB algorithm in our simulations. Therefore, the obtained results were compared with two recent computational studies: one for low translational Reynolds numbers \cite{wlf_low} $Re_{\gamma} = \{0.1, 0.01\}$ and other one for intermediate-low \cite{wlf_heigh} -- $Re_{\gamma} = \{1,3,10\}$;

Normally, the WLF is measured as a transverse force acting on a spherical particle moving in a channel parallel and close to the wall (see Fig. \ref{fig:low_wlf_plot}). In the LB scheme employed in this work, one needs to use a so-called raspberry particle \cite{raspberry} instead of a sphere in order to physically model the torques acting from the solvent on the particle. A raspberry particle is a sphere uniformly filled with a big amount of small beads (approximately 10-20 beads per unit cubic mesh) rigidly connected to the one central bead. One \textit{real} bead is located in the centre of mass and all other \textit{virtual} beads behave as tentacles that sense liquid and transfer all applied on them forces and torques to the real particle. 


\subsection*{Test 1: low Reynolds numbers}

\begin{figure}[h!]
	\centering
		 \begin{subfigure}[]
    		 {\label{fig:low_wlf_plot}
    		    \imagebox{45mm}{
    		        \includegraphics[width=0.19\textwidth]{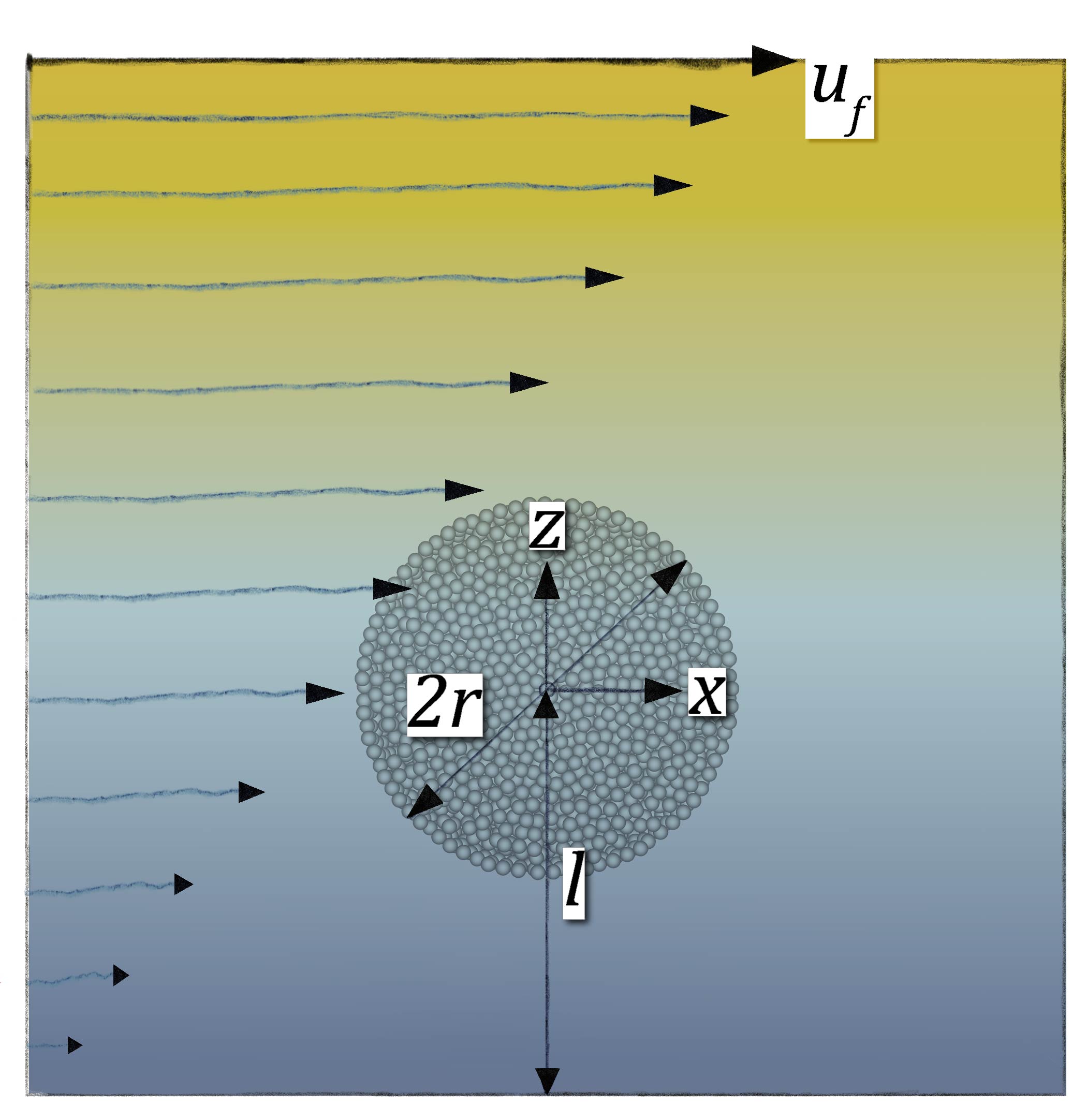}
    		    }
    		    
    		 }
		 \end{subfigure}
		 \begin{subfigure}[]
		    {\label{fig:low_wlf_raspberry}
		    \imagebox{45mm}{
		        \includegraphics[width=0.22\textwidth]{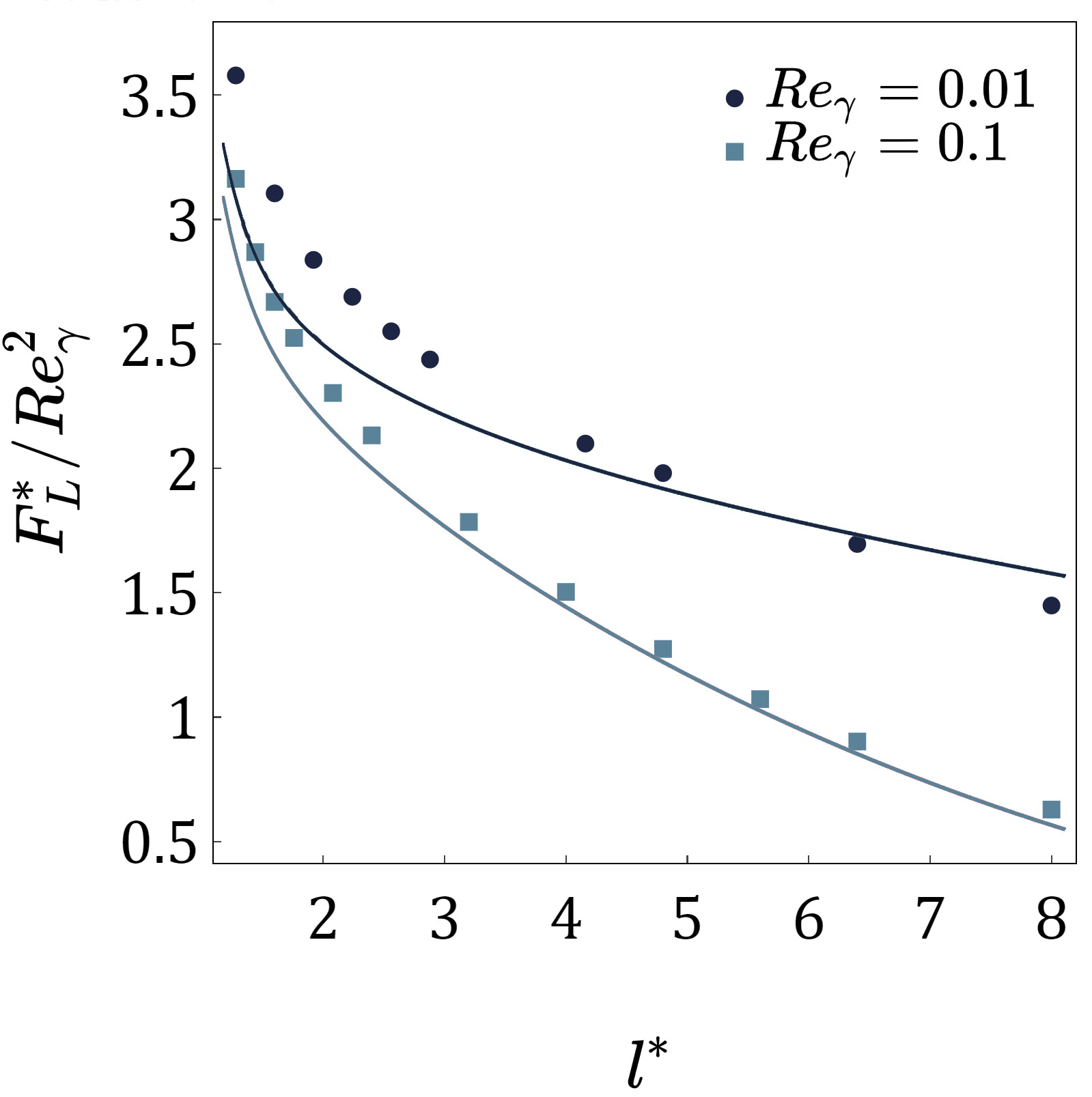}
		        }
		    }
		 \end{subfigure}
	\caption{(a) The sketch of the system in which the WLF (from the side of the bottom wall) was measured in the 1st test, as well as in the corresponding work \cite{wlf_low}. The upper wall moves at a speed of $u_{f}$, creating a shear flow, the lines of which are shown in the picture. $r$ is the radius of the raspberry-like particle (or sphere in \cite{wlf_low}). Distance between the center of the particle and the wall is $l$. The scale is distorted for visualization purposes. (b) The ration between WLF and the square of the $Re_{\gamma}$ ploted as a function of the dimensionless distance: $l^{*}=l/r$. Symbols are results of the simulations; solid lines is the date from \cite{wlf_low}.}
	\label{fig:low_wlf}
\end{figure}

In the work \cite{wlf_low}, the authors measured the lift and drag forces acting on a spherical neutrally-buoyant particle in a single wall-bounded linear shear flow field under a zero-slip condition, {\it i.e.} the particle translates with the same velocity as the fluid. This study is based on applying a numerical method to solve the steady–state Navier–Stokes equation utilising a finite volume solver over a non-uniform body-fitted structured mesh.  The sketch is provided in Fig. \ref{fig:low_wlf_plot}. Here, instead of a solid sphere, we directly sketch the raspberry particle that we used to test the parameters of our hydrodynamic scheme.  Thermal fluctuations are absent\footnote{$k_{b}T = 0$. The same is true for Test 2 below.}.  

We focused on the lift force $F_{L}^{*}$ acting from the bottom wall on the sphere along $z$-axis\footnote{In general, the lift force can arise without any wall due to the particle rotation (Magnus force) or non-zero slip velocity.}. More precisely, we study how the magnitude of this force depends on the distance between the center of the sphere and the bottom wall $l$ as shown in Fig. \ref{fig:low_wlf_raspberry} that corresponds to  Fig.~7(e) in the work of Ekanayake et al \cite{wlf_low}. Here, our results are presented with symbols, while the reference data is plotted with solid lines. As it can be seen, for $Re_{\gamma} = 0.1$, both approaches agree very well and for $Re_{\gamma} = 0.01$, at small $l^*$ the $F_{L}^{*}$ is slightly overestimated by the LB approach that can be attributed to the resolution of the mesh and the absence of a zero-slip condition in our simulations. Despite little discrepancies, Fig. \ref{fig:low_wlf_raspberry} allows to conclude that the LB fluid of a given viscosity set in our simulations is appropriate and physically meaningful for modelling low-Re flows. It is worth noting here that for a relatively fine mesh the compared system requires $\sim$ 1.2Tb of computational memory, while our coarse mesh fits in 4-5 Gb; nowadays, such a difference affects the feasibility of computer simulations in general.

\subsection*{Test 2: intermediate-low Reynolds numbers}

\begin{figure}[h!]
	\centering
		 \begin{subfigure}[]
    		 {\label{fig:high_wlf_plot}
    		    \imagebox{45mm}{
    		        \includegraphics[width=0.19\textwidth]{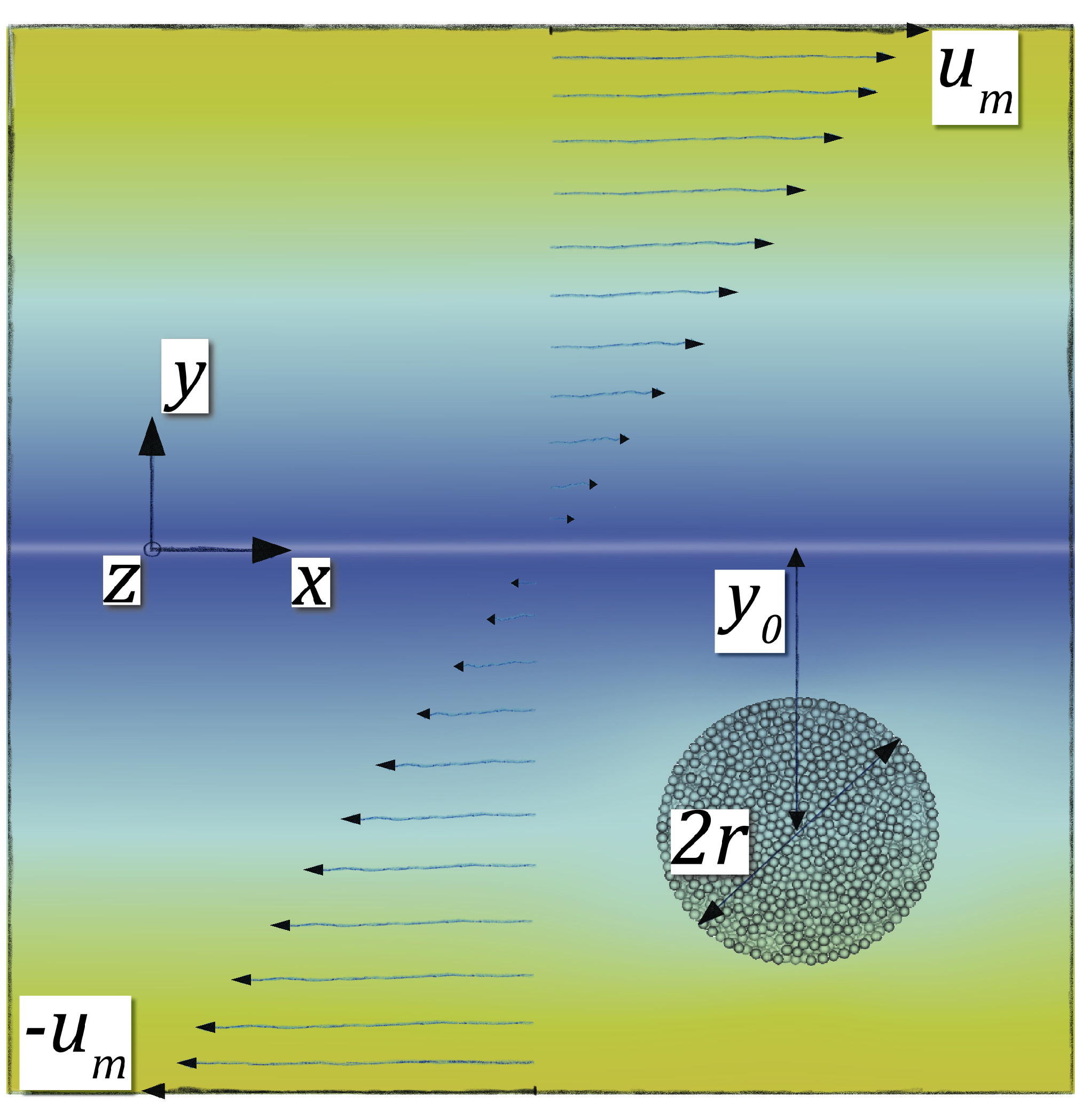}
    		    }
    		 }
		 \end{subfigure}
		 \begin{subfigure}[]
		    {\label{fig:high_re}
		    \imagebox{45mm}{
		        \includegraphics[width=0.22\textwidth]{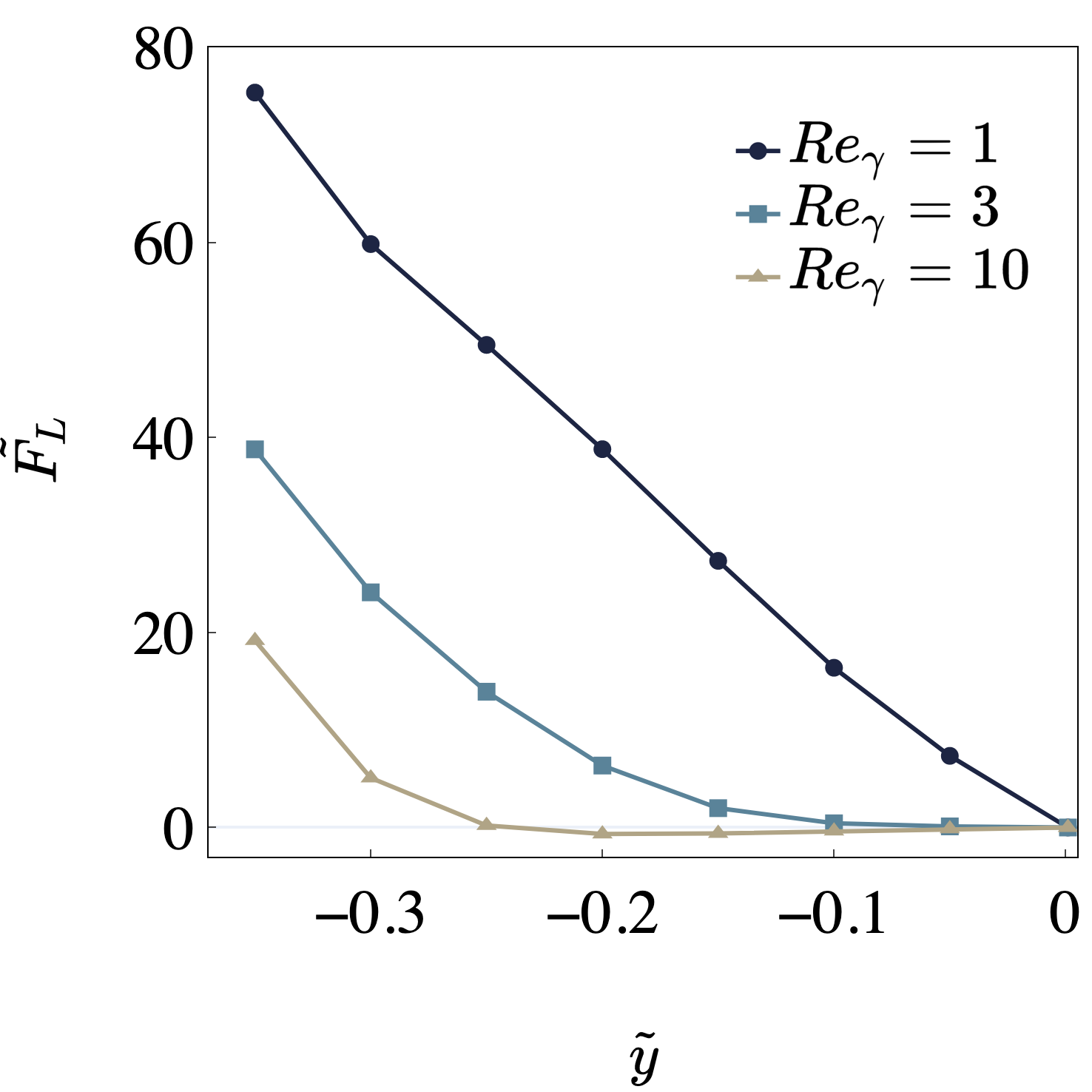}
		        }
		    }
		 \end{subfigure}
	\caption{(a) The sketch of the system in which the WLF ($F_{L}$, from the side of the bottom wall) was measured in the 2nd test, as well as in the corresponding work \cite{wlf_heigh}. Both, top and bottom walls move at the speed of $u_{m}$ in opposite directions, creating a shear flow, the lines of which are shown in the picture. $r$ is the radius of the raspberry-like particle (or of an infinite along $z$-axis cylinder in \cite{wlf_heigh}). $y_{0}$ is particles transverse position from channel center. The scale is preserved. (b) Scaled WLF (see eq.~\ref{eq:wlf_high}). $\tilde{y}=y_{0}/h$.  Based on symmetry considerations, WLF was measured only in the lower half of the channel. One can see a bifurcation point for $Re_{\gamma}=10$ at $\tilde{y}=-0.24$.}
	\label{fig:high_wlf}
\end{figure}

In the second test, we measured the same distance dependence of the lift force as in the first one, but for a slightly different geometry of the experiment. In this case, the top and bottom walls of the channel move in opposite directions as described in Fig.~\ref{fig:high_wlf_plot}. In the corresponding work of Fox et al \cite{wlf_heigh}, a home-made code implementing the LB algorithm in 2D was used, {\it i.e.} instead of a sphere, an infinite cylinder was modelled. We, in contrast, used the full 3D setup aiming at capturing the qualitative behaviour.

\vskip 0pt

\vskip 0pt
Indeed, as  Fig.~\ref{fig:high_re} show, our results plotted with symbols capture properly the trends calculated in the aforementioned article \cite{wlf_heigh} and presented there in Fig. 3. We believe that the quantitative discrepancy may be due to the fact that the normalisation of the WLF by the force per unit length in 2D as performed by Fox et al \cite{wlf_heigh}:

\begin{equation}
\tilde{F_{L}} = \frac{F_{L}}{\rho u_{m}^{2} \frac{r^{3}}{h^{2}}};
\label{eq:wlf_high}
\end{equation}
is not directly applicable for the 3D system investigated here. In Eq. \eqref{eq:wlf_high}, $h$ is the distance between bottom and top walls. Important validation for our approach, however, is that we reproduce the bifurcation of the equilibrium position, {\it i.e.} the new point near the wall where the $\tilde{F_{L}} = 0$  for $Re_{\gamma} = 10$, and it is this effect that the aforementioned article \cite{wlf_heigh} is devoted to in the first place. Thus, we can rely on our LB fluid parameters to study the flows not only with low Reinolds numbers, but also up to  $Re \lesssim 10$. This range is sufficient for the investigation of the MNG behaviour in the shear flow presented in this work.

\section*{Appendix B: Magnetic clustering}\label{sec:app-B}
\begin{figure}[h!]
	\centering
	\includegraphics[width=0.23\textwidth]{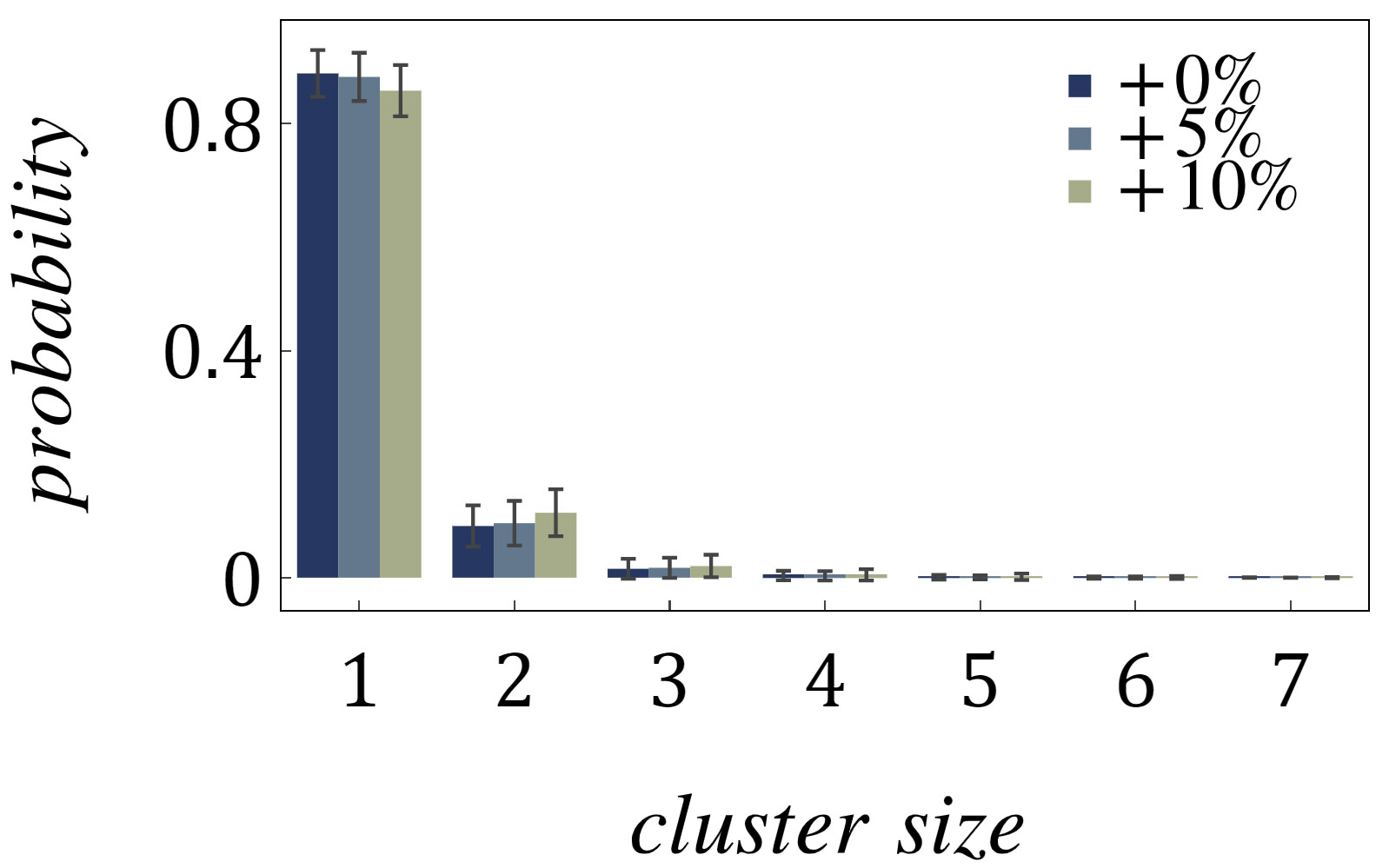}
	\caption{The histogram shows the probability to find a MNPs cluster of a given size for different values of their volume fraction relative to default 10 per cent taken in this study, as shown in the legend. $\lambda = 5$}
	\label{fig:mnps_distr_with_add_mnps}
\end{figure}

In order to confirm that periodic increase of the cluster size observed in a MNG is not simply a statistical error, but is rather the consequence of the MNP concentration increase by approximately 10 per cent due to the oscillations of the MNG volume,  we make the following experiment. We simulate a MNG in the channel with no shear,  $\dot{\gamma}=0$,  changing the number of MNPs inside it so that the volume fraction is increased either by 5 or by 10 per cent. The value of $\lambda$ is fixed, $\lambda = 5$,  to achieve this value, particles of cobalt ferrite with a slightly larger core size were taken. In  Fig. \ref{fig:mnps_distr_with_add_mnps}, we plot the histogram of the cluster size distributions for different concentrations as indicated in the legend. One can clearly observe the same effect as previously shown in Fig. \ref{fig:clusters-hist-l3}. 

\bibliographystyle{elsarticle-num-names}

\bibliography{gels2021.bib, Mendeley.bib}

\end{document}